\documentclass[11pt,journal,onecolumn]{IEEEtran}
\usepackage{epsfig,setspace,amsmath,epsf,amssymb,bm,theorem,cite,graphicx, epstopdf,algorithm,algpseudocode,float,mathtools,authblk,tikz,physics}

\usetikzlibrary{positioning}
\usetikzlibrary{decorations.text}
\usetikzlibrary{decorations.pathmorphing}
\usepackage[short,c2]{optidef}

\usepackage{empheq}
\usepackage{cases}

\usepackage{booktabs}
\usepackage{subfigure}
\usepackage{bbm}
\usepackage{amsmath,amssymb}
\usetikzlibrary{arrows.meta,positioning,calc,shapes.geometric}
\newtheorem{theorem}{Theorem}

\newtheorem{remark}{Remark}
\newtheorem{lemma}{Lemma}

\newenvironment{Proof}[1]{\medskip\par\noindent{\bf Proof:\,}\,#1}{{\mbox{\,$\blacksquare$}\par}}

\usepackage[normalem]{ulem}
\usepackage{setspace}
\IEEEoverridecommandlockouts
\allowdisplaybreaks
\begin{document}

\title{Elapsed-Service-Time-Aware Preemption for Age of Information Minimization
}

\author{Mohammad~Moltafet,~\IEEEmembership{Member,~IEEE,}
        Nail~Akar,~\IEEEmembership{Senior~Member,~IEEE,}
        Sennur~Ulukus,~\IEEEmembership{Fellow,~IEEE,} \\
        Marian~Codreanu,~\IEEEmembership{Member,~IEEE,}
        and~Roy~D.~Yates,~\IEEEmembership{Fellow,~IEEE}
\thanks{M. Moltafet is with the Department
of Electrical and Computer Engineering, Tennessee Tech University, TN, USA
 (e-mail: mmoltafet@tntech.edu), N. Akar is with Bilkent University, Ankara, T\"{u}rkiye (email: akar@ee.bilkent.edu.tr), S. Ulukus is with University of Maryland, College Park, USA (email: ulukus@umd.edu), M. Codreanu is with Link\"oping University, Link\"oping, Sweden  (email: marian.codreanu@liu.se), and R. D. Yates is with Rutgers University, New Brunswick, NJ, USA  (email: ryates@winlab.rutgers.edu).}
 \thanks{A preliminary version of this work was presented
at the IEEE MILCOM 2026 Workshop on Quality, Value,
and Age of Information for Tactical Networks~\cite{Squ_Most_026}.}
}

\maketitle

\begin{abstract}
We study the age of information and peak age of information
in status update systems with an arbitrary number of
independent sources sharing a single server with generally
distributed service times. We consider a general
age-dependent self-preemptive policy in which an in-service
update from source $c$ is preempted by a new same-source
arrival with a probability $p_c(x)$ that depends on the
elapsed service time $x$. This formulation subsumes the
non-preemptive, fully self-preemptive, and probabilistically
preemptive policies as special cases, and also accommodates
threshold-based preemption. We develop a unified
analytical framework based on non-homogeneous absorbing
continuous-time Markov chains and semi-Markov chains to
derive explicit expressions for the average age of information
and peak age of information of each source under general
per-source service-time distributions and age-dependent
preemption rules.
The framework applies to both single-source and
multi-source systems and enables joint tuning of
per-source preemption parameters. Numerical results
for fifteen service-time distributions in the single-source
setting reveal that the benefits of preemption depend
strongly on the service-time hazard-rate structure,
with threshold-based preemption providing its most notable
additional gains over age-independent policies for
distributions with unimodal hazard rates. For a dual-source
system with lognormal service times, jointly tuned
threshold-based preemption achieves lower sum average
age of information and peak age of information than the
considered age-independent policies, highlighting the
importance of cross-source interactions in preemption
decisions.
\end{abstract}
\begin{IEEEkeywords}
Age of information, continuous-time Markov chains, packet
management, peak age of information, preemption.
\end{IEEEkeywords}

\section{Introduction}\label{Introduction}
Timely delivery of status updates is a key requirement in time-critical applications such as remote monitoring, networked
control systems, and the Internet of Things (IoT)
\cite{elmagid_commag19,abbas_survey23}.
The age of information (AoI) was first characterized in \cite{5984917}
as a destination-centric metric that quantifies
information freshness by measuring the elapsed time since the most
recently received status update was generated at the source.
Unlike delay-based metrics, AoI captures both the timeliness and
the regularity of updates, making it particularly suited for
evaluating the performance of status update systems
\cite{kosta_etal_survey,yates_survey}.
The peak AoI (PAoI), proposed in~\cite{costa_peak}, measures the AoI
immediately before a new update is successfully received and
serves as an alternative freshness metric that is often more
tractable analytically.

\subsection{Background and Related Work}

Age of information was first studied from a queueing-theoretic standpoint in \cite{Yates__HowOftenShouldone}, where the average AoI was derived for single-source M/M/1, M/D/1, and D/M/1 queues under first-come-first-served (FCFS) scheduling. 
The work in \cite{6284003} was the first to consider AoI in a multi-source setting, analyzing the average AoI of a multi-source M/M/1 FCFS queue.
It has since been shown that an appropriate packet management policy
can substantially improve information freshness.  The average AoI of an M/M/1 last-come-first-served (LCFS) queue with preemption in service was derived in \cite{6310931}. The work in \cite{costa_etal_TIT16} investigates the AoI and PAoI for the bufferless M/M/1/1 queue and the single-buffer M/M/1/2 queue, along with the M/M/1/2$^{\ast}$ model in which a packet waiting in the queue is replaced by a fresh arrival.

Beyond direct queueing-theoretic derivations, a large part of the AoI literature relies on the stochastic hybrid systems (SHS) technique, introduced in \cite{yates2019} to compute the average AoI by tracking an auxiliary age process through linear differential equations between reset events, and later extended in \cite{yates2020age} to obtain the moment generating function (MGF) of the AoI; a comprehensive treatment of the SHS technique and of AoI research more broadly is given in the survey \cite{yates_survey}. The SHS technique has since been used to analyze a wide range of preemptive queueing models.  In \cite{8406966}, the average AoI is derived for a single-source system whose updates traverse a series of network nodes, each employing an LCFS queue that supports preemption in service. The work in \cite{yates_isit18} derives the average AoI for a single-source system with multiple parallel servers, each supporting preemption in service. In \cite{8437591}, the authors consider a multi-source system in which sources have different priorities and derive the average AoI under two priority-based packet management policies. The work in \cite{javani_etal_globecom19} considers a multi-source LCFS system with multiple servers under a \emph{globally preemptive} policy, where packets of different sources can preempt one another. The work in \cite{9048914} derives the average AoI for a multi-source system under a \emph{self-preemptive} policy with packet delivery errors, where an arriving packet only preempts an in-service packet from the \emph{same} source. The works \cite{9252168,9162681} derive the average AoI for a multi-source M/M/1 queue under several preemptive and non-preemptive packet management policies, \cite{Moltafet2020mgf} derives the MGF of the AoI for a multi-source M/M/1 queue under various packet management policies, and \cite{9611498} derives the MGF of the AoI for a two-server tandem queue in which status updates require additional processing before being used. SHS analysis is most naturally suited to Markovian (exponential) service and arrival processes, since the underlying age process must then evolve according to simple linear dynamics between resets.

Besides exponential service and Poisson arrivals, AoI has also been studied under more general arrival and service processes, in both single-source and multi-source systems; the analytical method developed in this paper falls within this latter strand. In the single-source setting, the average AoI and PAoI for the preemptive LCFS M/G/1/1 queue are derived in \cite{najm_nasser_isit16}, where the service time is modeled by the gamma distribution, while \cite{akar_etal_tcom20} investigates the distributions of the AoI and PAoI in single-source bufferless systems with probabilistic preemption, allowing phase-type distributions for both interarrival and service times. The work in \cite{8406909} analyzes the AoI in a single-source D/G/1 FCFS queue, and \cite{8006504} derives a closed-form average AoI expression for a single-source M/G/1/1 preemptive queue with hybrid automatic repeat request. \cite{inoue_etal_IT19} presents a general formula for the steady-state distributions of the AoI and PAoI for a broad class of single-source single-server queues. The work in \cite{soysal_ulukus_IT21} derives upper bounds for the mean AoI of the G/G/1/1 queue as well as its preemptive version, showing that the bounds are close to the actual values. The work in \cite{moltafet_etal_isit25} studies a preemptive server with generally distributed service times under a fixed, age-independent preemption probability and presents the advantages of probabilistic preemption for certain service time distributions. Preliminary results of the present work were reported in \cite{Squ_Most_026}, where age-dependent preemption probabilities were analyzed for a single-source system using non-homogeneous absorbing continuous-time Markov chains (CTMCs), and probabilistic and threshold-based policies were evaluated for lognormal service times. The present paper extends this analysis to an arbitrary number of sources and substantially broadens the numerical investigation of the single-source case. More recently, \cite{li_2026_tamin} considers a single-source generate-at-will status update system with generally distributed service times and studies a continuous-time joint sampling and preemption problem to minimize the AoI, while \cite{banerjee_ulukus_isit24} explicitly studies the question of when to preempt in a status update system. Our age-dependent preemptive policy can be viewed as a tractable, analytically tunable instance of this broader design question.

In the multi-source setting with generally distributed service times, \cite{7282742} optimizes the per-source arrival rates of a multi-source M/G/1 queue to minimize the peak AoI. The work of \cite{najm_telatar_infocom18} investigates an M/G/1/1 queue under a globally preemptive policy, where an arriving packet preempts the in-service packet irrespective of source index, and computes a closed-form expression for the average AoI and PAoI of each source by using the detour flow graph. The average AoI and PAoI under a non-preemptive policy, where any arrival is discarded while the server is busy, are derived for a multi-source M/G/1/1 queue in \cite{9500775}. In \cite{8886357}, the authors derive the average AoI for a system with two classes of Poisson arrivals and general service times, under different priorities, where a newly arriving packet replaces an in-service packet of the same or lower priority. When service time distributions are phase-type, the authors of \cite{dogan_akar_tcom21} provide an algorithm to obtain the distributions of the AoI and PAoI for a multi-source M/PH/1/1 model under a probabilistically preemptive policy, in which a packet from source $c$ preempts an in-service packet from source $c'$ with a probability depending on the pair $(c,c')$; the self-preemptive, globally preemptive, and non-preemptive policies all arise as special cases of this framework. The work in \cite{moltafet_etal_tcom22} investigates a multi-source M/G/1/1 queueing model under a self-preemptive packet management policy and derives the MGF of the AoI and PAoI, and this self-preemptive setting is further generalized in the work \cite{Moltafet_etal_Asilomar25}, where 
an in-service packet is preempted by a same-source arrival with a fixed probability $\alpha_c$. Other relevant multi-source studies include the FCFS M/G/1 approximations in \cite{moltafet2019age}, its exact extension in \cite{inoue2024exact}, and the preemptive packet management analysis in \cite{10038591}. In addition, 
the work in
\cite{akar_gamgam_comlet23} uses an absorbing Markov chain formulation to obtain the AoI distribution for a multi-source system with heterogeneous information sources. A recent related work \cite{10994455} considers a multi-class system with Erlang service times and employs a hybrid service discipline in which scheduling decisions depend on the elapsed service phase. In contrast, our framework allows generally distributed service times and makes the self-preemption decision directly as a function of the continuous elapsed service time.

Across this body of work, age-dependent preemption 
has received limited attention.

\subsection{Motivation and Contributions}
As established above, existing studies of preemptive status update systems generally adopt some form of age-independent preemption rule, while age-dependent preemption with general service times has been investigated primarily in the single-source setting, including our preliminary work in \cite{Squ_Most_026}.
In many practical deployments, e.g.,  sensor networks, vehicular telemetry, and industrial monitoring, several independent sources of differing importance, generation rates, and service requirements compete for access to a common communication or processing resource. 
A natural and largely unexplored question is  whether, and how, each source's preemption behavior should depend on the elapsed service time of its own update in service, and how such individually tuned rules interact once several sources share the same server. 

We consider a status update system with an arbitrary number of independent sources sharing a single server, including the single-source case as a special case.
We assume that each source generates updates according to a Poisson process and the  service time for an update  is generally distributed. We consider a general age-dependent self-preemptive policy.
According to the policy, when a new update from source $c$ arrives, it immediately enters service if the server is idle.
If a new update from source $c$ arrives while the server is serving an update from the same source, the arriving update preempts the in-service update with probability $p_c(x)$, where $x$ is its elapsed service time. Intuitively, an update that has spent an unusually long time in service may be worth replacing with a fresher update, whereas preempting a recently started update may unnecessarily waste service effort.

Next, we summarize our contributions. 

\begin{itemize}
    \item We extend the analytical approach of our preliminary single-source
    study~\cite{Squ_Most_026} to an arbitrary number of sources. The resulting
    framework, based on non-homogeneous absorbing CTMCs and semi-Markov
    chains, yields explicit analytical expressions for the average AoI
    and PAoI of every source under general per-source service-time
    distributions and age-dependent self-preemption probabilities.
    The non-preemptive (NP), fully self-preemptive (FP), and
    probabilistically preemptive (PP) policies arise as special cases.

    \item We substantially expand the single-source numerical investigation
    to fifteen service-time distributions and examine how the shape of the
    service-time hazard rate affects the performance of preemption policies.
    In particular, threshold-based preemption (TP) provides its most
    notable additional gains over age-independent policies for
    distributions with unimodal hazard rates.

    \item We propose a two-stage joint tuning procedure that combines a
    coarse grid search with Nelder--Mead refinement and exploits
    the derived expressions to numerically tune the per-source
    parameters of both the PP and TP policies. Numerical results for a
    dual-source system with lognormal service times show that the jointly
    tuned TP policy achieves lower sum average AoI and PAoI than the
    considered age-independent policies and reveal a cross-source
    interaction through the shared server, highlighting the importance
    of joint per-source tuning.
\end{itemize}




The main extension beyond our preliminary single-source results~\cite{Squ_Most_026} is the characterization of age-dependent self-preemption for an arbitrary number of sources with generally distributed service times. This extension captures interactions through the shared server and enables joint tuning of per-source preemption parameters, while the expanded single-source experiments clarify the role of service-time hazard-rate structure.
\subsection{Organization}
The remainder of this paper is organized as follows. Section~\ref{system_sec} describes the system model, defines the packet management policy, and states the main results. Section~\ref{analytical_sec} presents the derivation of the average AoI and PAoI, consisting of the lifetime analysis via non-homogeneous absorbing CTMCs and the inter-delivery time characterization via a semi-Markov chain. 
Section~\ref{results_sec} investigates the single-source case across fifteen service-time distributions, with particular attention to hazard-rate characteristics. Section~\ref{Multi-source} examines the dual-source case and the joint tuning of preemption parameters.
Section~\ref{conclusions_sec} concludes the paper.

\section{System Model and 
Main Results}\label{system_sec}

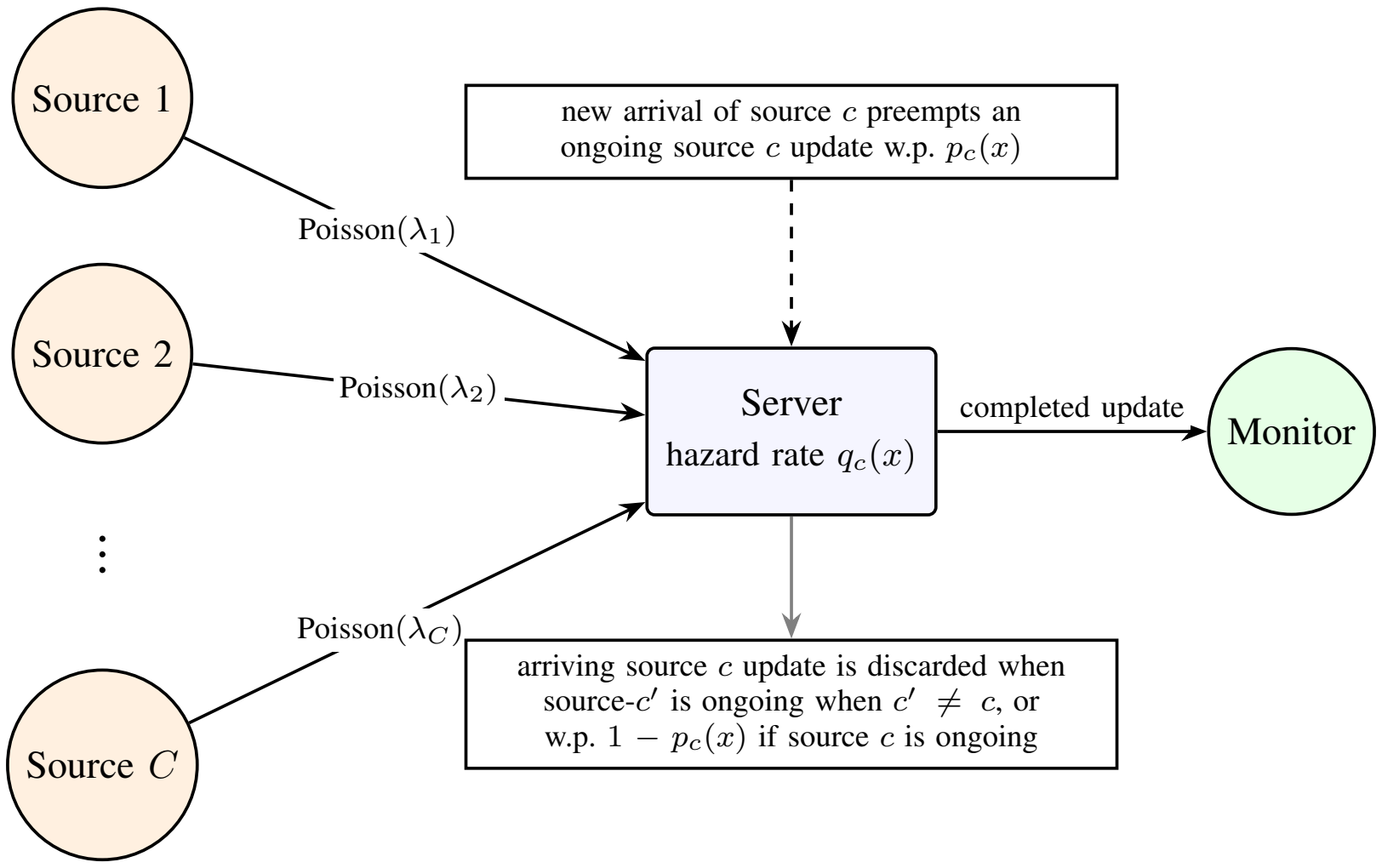
\begin{figure}
\centering
\resizebox{0.6\textwidth}{!}{
\begin{tikzpicture}[
    >=Stealth,
    thick,
    font=\small,
    block/.style = {draw, rounded corners=2pt, minimum width=2.6cm,
                     minimum height=1.5cm, align=center, fill=blue!4},
    src/.style   = {draw, circle, minimum size=1.1cm, align=center, fill=orange!12},
    dst/.style   = {draw, circle, minimum size=1.1cm, align=center, fill=green!10},
    dec/.style   = {draw, align=center, font=\scriptsize, fill=white},
    lbl/.style   = {font=\scriptsize, fill=white, inner sep=1.5pt},
]

\node[src] (src1) at (0, 3.4)  {Source 1};
\node[src] (src2) at (0, 1.1)  {Source 2};
\node[font=\Large] (dots) at (0, -0.6) {$\vdots$};
\node[src] (srcC) at (0, -2.6) {Source $C$};

\node[block] (srv) at (6.2, 0.4) {Server\\[1pt] \footnotesize hazard rate $q_c(x)$};
\node[dst]   (dst) at (10.7, 0.4) {Monitor};

\draw[->] (src1) -- node[lbl, pos=0.42] {Poisson$(\lambda_1)$} (srv);
\draw[->] (src2) -- node[lbl, pos=0.5]  {Poisson$(\lambda_2)$} (srv);
\draw[->] (srcC) -- node[lbl, pos=0.42] {Poisson$(\lambda_C)$} (srv);

\draw[->] (srv) -- node[lbl, above]{completed update} (dst);

\node[dec, above=1.5cm of srv.north, text width=5.6cm] (rule)
      {new arrival of source $c$ preempts an ongoing source $c$ update
      w.p. $p_c(x)$};

\draw[->, dashed] (rule.south) -- (srv.north);

\node[dec, below=1.1cm of srv.south, text width=5.6cm, text=black] (disc)
      {arriving source $c$ update is discarded when source-$c'$ is ongoing when $c' \neq c$, or w.p. $1-p_c(x)$ if source $c$ is ongoing} ;
\draw[->, gray] (srv.south) -- (disc.north);

\end{tikzpicture}
}
\caption{The multi-source self-preemptive status update system for which source $c$ generates updates according
to a Poisson process with rate $\lambda_c$. If a new source $c$ update arrives
while another source $c$ update is in service, it preempts the in-service update
with probability $p_c(x)$, where $x$ is the elapsed service time. Otherwise, the
new arrival is discarded. Cross preemption between two different sources is not allowed.}
\label{Model}		
\end{figure}

We consider a status update system consisting of a set of independent sources denoted by $\mathcal{C}=\{1,\ldots,C\}$ and a single server as shown in Fig.~\ref{Model}.
We assume that source $c\in\mathcal{C}$ generates updates according to a Poisson process with rate $\lambda_c$. 
Since updates of each source are generated according to a Poisson process and the sources are independent, the update generation in the system follows
the Poisson process with total arrival rate $\lambda = \sum_{c=1}^{C} \lambda_c$.
The service time for an update from source $c$, denoted by $S_c$, is generally distributed with probability density function (pdf) $f_c(x)$, cumulative distribution function (cdf) $F_c(x)$, and hazard rate
\begin{align}
    q_c(x) & = \lim_{h \rightarrow 0} \frac{1}{h} \Pr (x < S_c \leq x + h \,|\, S_c > x) \\&\nonumber= \frac{f_c(x)}{1-F_c(x)},
\end{align}
which is also known as the instantaneous conditional completion rate when the elapsed service time (update age) has the value $x$. For the hazard rates 
of well-known distributions and also their definitions, we refer the reader to \cite{book_oconnor,johnson1994continuous}.

\subsection{Packet Management Policy}
When a new update message from source $c$ arrives, it immediately enters service if the server is idle.
If a new update from source $c$ arrives while the server is serving an update from the same source, the arriving update preempts the in-service update with probability $p_c(x)$, where $x$ is its elapsed
service time. An arriving update from source $i \neq c$ does \emph{not} preempt an update from source $c$ currently in service; it is discarded upon arrival.
This \textit{self-preemptive}  model allows each source to replace its own stale updates, while arrivals from other sources cannot preempt an update currently in service.

The described general self-preemptive policy 
contains several well-known policies as special cases.
When $p_c(x) = 1$ for all $x \geq 0$, source $c$ applies a fully self-preemptive (FP) policy: every new arrival from source $c$ always preempts the ongoing update from the same source \cite{dogan_akar_tcom21}.
When $p_c(x) = 0$ for all $x \geq 0$, source $c$ applies a non-preemptive (NP) policy: arriving updates from source $c$ are discarded if source $c$ is already being served, and do not interfere with the ongoing service \cite{dogan_akar_tcom21}.
The probabilistically preemptive (PP) policy of \cite{Moltafet_etal_Asilomar25}  uses a fixed probability $\alpha_c \in [0,1]$, i.e., $p_c(x) = \alpha_c$ for all $x$,
%
%
with NP and FP recovered as the special cases $\alpha_c=0$ and $\alpha_c=1$, respectively.
In this paper, we propose the threshold-based preemptive (TP) policy, for which
\begin{align}
    p_c(x) = \begin{cases} 1 & x > \tau_c, \\ 0 & x \leq \tau_c, \end{cases}
\end{align}
for a given threshold $\tau_c \geq 0$.
The motivation behind the TP policy is the following: when the update age is small, i.e., the update has not been in service for long, it makes sense to let it continue service, since it may be closer to completion.
%
%
However, when an update has been in service for a long time, the information it carries may have already become outdated. Moreover, depending on the hazard rate of the service-time distribution, a long elapsed service time may indicate that a substantial amount of service time remains. In such a case, preempting the update with a fresh one may be advantageous, as it offers an opportunity to deliver more recent information, even though some service work is wasted.

\subsection{AoI and PAoI}
\label{sec:AoI}
For source $c \in\mathcal{C}$, let $G_{c,k}$ denote the generation time of the $k^{\text{th}}$  delivered update.  From here onward, we refer to the 
$k^{\text{th}}$ delivered update from source $c$ simply as ``update $c,k$''. Let $R_{c,k}$ denote the delivery time of update $c,k$, and $U_{c,k}$ denote the system time (sojourn time) of update $c,k$, i.e., $U_{c,k} = R_{c,k} - G_{c,k}$.
At a time instant $\tau$,  the  index of the most recently received update of source $c$ is given by
\begin{equation}\label{mnb00}
N_c(\tau)=\max\{i'\mid R_{c,i'}\le \tau\}.
\end{equation}
The AoI of source $c$ at the sink is defined as the random process
\begin{align}  
\Delta_{c}(t)=t-G_{c,N_c(t)}.
\end{align}

\begin{figure}[tb]
    \centering
    \begin{tikzpicture}[scale=0.3]

    \draw[thick,->] (1,0) -- (32,0) node[anchor=north] {$t$};
    \draw[thick,->] (1,0) -- (1,16) node[anchor=south] {$\Delta_c(t)$};

    \filldraw[blue] (2,2)     circle (3pt);
    \filldraw[blue] (2.8,2.8) circle (3pt);
    \filldraw[blue] (3.6,3.6) circle (3pt);

    \draw[ultra thick,blue] (4.5,4.5) -- (9,9);
    \draw[ultra thick,blue] (9,9)   -- (9,3);
    \draw[ultra thick,blue] (9,3)   -- (17,11);
    \draw[ultra thick,blue] (17,11) -- (17,6);
    \draw[ultra thick,blue] (17,6)  -- (25,14);
    \draw[ultra thick,blue] (25,14) -- (25,4);
    \draw[ultra thick,blue] (25,4)  -- (29,8);

    \filldraw[blue] (29.5,8.5) circle (3pt);
    \filldraw[blue] (30.2,9.2) circle (3pt);
    \filldraw[blue] (30.9,9.9) circle (3pt);

    \draw[dotted,gray,thick] (6,0)  -- (9,3);
    \draw[dotted,gray,thick] (11,0) -- (17,6);
    \draw[dotted,gray,thick] (21,0) -- (25,4);

    \draw[dotted,gray,thick] (9,0)  -- (9,9);
    \draw[dotted,gray,thick] (17,0) -- (17,11);
    \draw[dotted,gray,thick] (25,0) -- (25,14);

    \node[anchor=north,black] at (6,0)     {\scriptsize $G_{c,k}$};
    \node[anchor=north,black] at (9,0)     {\scriptsize $R_{c,k}$};
    \node[anchor=north,black] at (11,-1.5) {\scriptsize $G_{c,k+1}$};
    \node[anchor=north,black] at (17,0)    {\scriptsize $R_{c,k+1}$};
    \node[anchor=north,black] at (21,-1.5) {\scriptsize $G_{c,k+2}$};
    \node[anchor=north,black] at (25,0)    {\scriptsize $R_{c,k+2}$};

    \draw[dotted,gray] (1,9)  -- (32,9);
    \draw[dotted,gray] (1,11) -- (32,11);
    \draw[dotted,gray] (1,14) -- (32,14);

    \draw (1,9)  node[anchor=east] {\scriptsize $\Phi_{c,k}$};
    \draw (1,11) node[anchor=east] {\scriptsize $\Phi_{c,k+1}$};
    \draw (1,14) node[anchor=east] {\scriptsize $\Phi_{c,k+2}$};

    \draw[dotted,gray] (1,3) -- (32,3);
    \draw[dotted,gray] (1,6) -- (32,6);
    \draw[dotted,gray] (1,4) -- (32,4);

    \draw (1,3) node[anchor=east] {\scriptsize $U_{c,k}$};
    \draw (1,6) node[anchor=east] {\scriptsize $U_{c,k+1}$};
    \draw (1,4) node[anchor=east] {\scriptsize $U_{c,k+2}$};

    \draw[<->,black,thick] (9,15.5)  -- (17,15.5);
    \node[anchor=south] at (13,15.5) {\small $V_{c,k}$};

    \draw[<->,black,thick] (17,15.5) -- (25,15.5);
    \node[anchor=south] at (21,15.5) {\small $V_{c,k+1}$};

    \draw[gray] (9,3)  circle (6pt);
    \draw[gray] (17,6) circle (6pt);
    \draw[gray] (25,4) circle (6pt);

    \end{tikzpicture}
    \vspace{-2mm}
    \caption{Sample path of the AoI process $\Delta_c(t)$. Only generation and reception instances of successfully delivered packets are shown.}
    \label{fig:samplepath}
    \vspace{-2mm}
\end{figure}
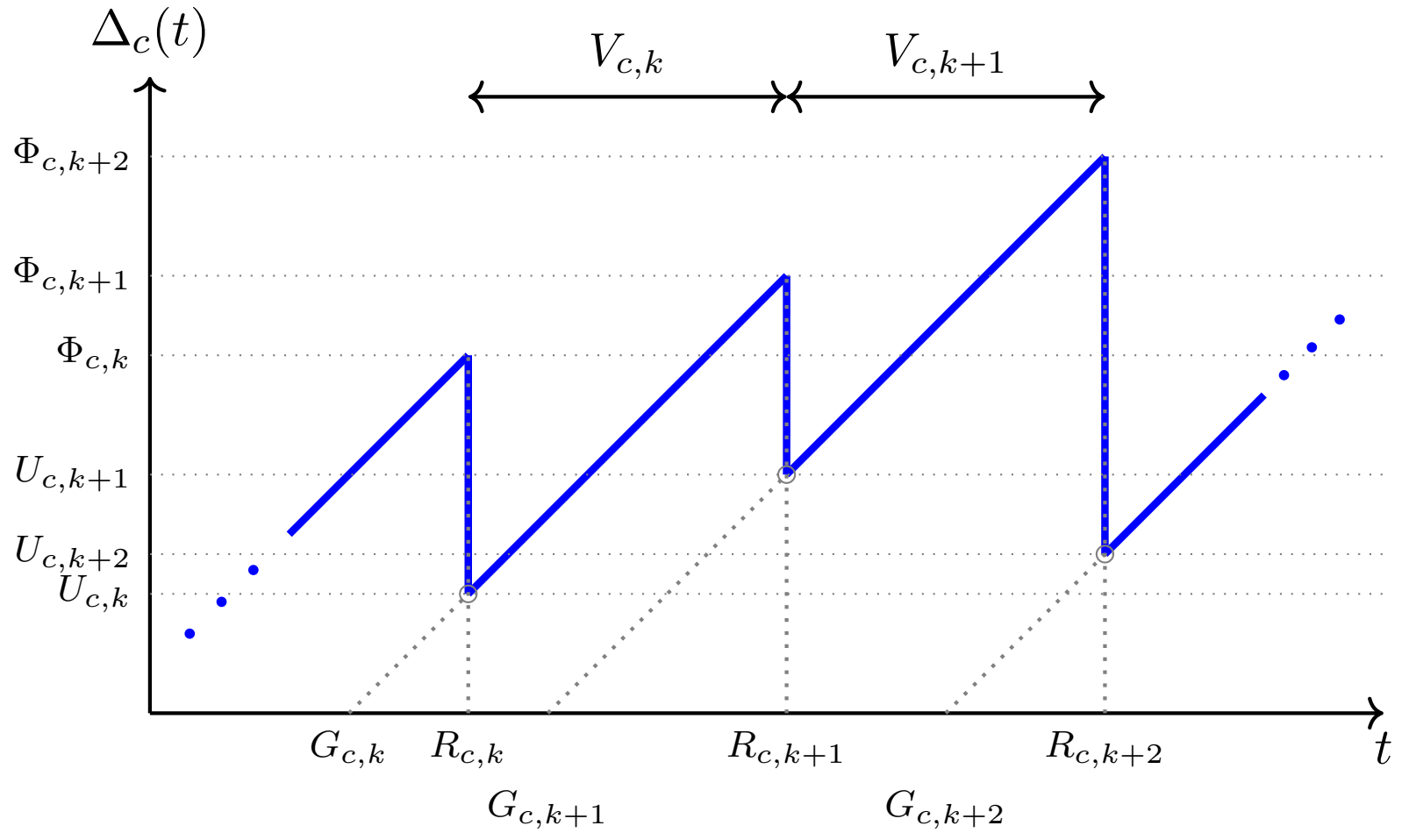


A sample path of the AoI process $\Delta_c(t)$ is illustrated
in Fig.~\ref{fig:samplepath}. The AoI process $\Delta_c(t)$
for source $c$ increases with unit slope between two
consecutive reception instances $R_{c,k}$ and $R_{c,k+1}$,
starting from the value $U_{c,k}$ at time $R_{c,k}$ and
reaching
\begin{align}
    \Phi_{c,k+1}=U_{c,k}+V_{c,k},
\end{align}
just before time $R_{c,k+1}$, where
$V_{c,k}=R_{c,k+1}-R_{c,k}$ is the $k$th inter-delivery
time of source $c$. At time $R_{c,k+1}$, the AoI drops to
$U_{c,k+1}$, initiating a new cycle. The PAoI sequence
$\{\Phi_{c,k}\}$ is obtained by sampling $\Delta_c(t)$ just
before the reception epochs $R_{c,k}$.

We assume that the considered status update system is
stationary and that the AoI process for each source is ergodic. Let  $\Delta_c$ and $\Phi_c$ denote the steady-state random variables corresponding to the AoI and PAoI, respectively. Then, the mean AoI and mean PAoI can be characterized as \cite{yates_survey}
\begin{align}
  \mathbb{E}[\Delta_c]  & = \dfrac{1}{\mathbb{E}[V_{c,k}]}\left(\mathbb{E}[V_{c,k}U_{c,k}] + \frac{\mathbb{E}[V_{c,k}^2]}{2 }\right), \label{meanAoIexp}\\
    \mathbb{E}[\Phi_c] & = \mathbb{E}[U_{c,k}] + \mathbb{E}[V_{c,k}]. \label{meanPAoIexp0}
\end{align}
Under the considered packet management policy, every
successful delivery leaves the server idle. By the independent
increments of the Poisson arrival processes and the
independence of service times, the system time $U_{c,k}$ of the
delivered update is independent of the subsequent
inter-delivery time $V_{c,k}$.
Thus, \eqref{meanAoIexp} simplifies to
\begin{align}
  \mathbb{E}[\Delta_c]  & = \mathbb{E}[U_{c,k}] + \frac{\mathbb{E}[V_{c,k}^2]}{2 \mathbb{E}[V_{c,k}]}. \label{meanAoIexp_O0}
\end{align}

For notational simplicity, let $U_c$ and $V_c$ denote random
variables having the steady-state distributions of $U_{c,k}$
and $V_{c,k}$, respectively, i.e., $U_c \sim U_{c,k}$ and
$V_c \sim V_{c,k}$, where $\sim$ denotes equality in
distribution. Then, the mean AoI and mean PAoI are given by
\begin{align}
  \mathbb{E}[\Delta_c]  & = \mathbb{E}[U_{c}] + \frac{\mathbb{E}[V_{c}^2]}{2 \mathbb{E}[V_{c}]}, \label{meanAoIexp_O}\\
    \mathbb{E}[\Phi_c] & = \mathbb{E}[U_{c}] + \mathbb{E}[V_{c}]. \label{meanPAoIexp}
\end{align}

The expressions in \eqref{meanAoIexp_O} and \eqref{meanPAoIexp} show that to compute the average AoI and PAoI for source $c$, it suffices to obtain $\mathbb{E}[U_c]$, $\mathbb{E}[V_c]$, and $\mathbb{E}[V_c^2]$.
The analytical method proposed in this paper provides a systematic way to compute these three quantities for the multi-source self-preemptive system.

Next, we derive the average AoI $\mathbb{E}[\Delta_c]$ and average PAoI $\mathbb{E}[\Phi_c]$ for each source $c \in \mathcal{C}$ in the considered multi-source update system, characterized by the $C$-tuple of 3-tuples $\{(\lambda_c, q_c(x), p_c(x))\}_{c=1}^C$.  

The derived expressions can subsequently be used to jointly
tune the per-source preemption parameters for a chosen
system-wide freshness objective, as illustrated in
Section~\ref{Multi-source}.

\subsection{Summary of the Main Results}
\label{sec:summary}
The average PAoI and average AoI of source $c$ in the considered status update system are given by the following two theorems. The proofs of the theorems are presented in Section~\ref{analytical_sec}.
For source $c$, define the lifetime of an update as the time it
spends in service until it is either preempted or successfully
completes service. Let $T_{p_c}$ and $T_{q_c}$ denote this
lifetime conditioned on preemption and successful completion,
respectively, and let $\varepsilon_c^p$ and $\varepsilon_c^q$
denote the corresponding probabilities of preemption and
successful completion, respectively.
\begin{theorem}[Average PAoI]
\label{thm:paoi}
Consider source $c \in \mathcal{C}$ in the multi-source self-preemptive system characterized by $\{(\lambda_c, q_c(x), p_c(x))\}_{c=1}^C$. The average PAoI of source $c$ is given by
\begin{align}
   \mathbb{E}[\Phi_c]= \mathbb{E}[T_{q_c}] + \mathbb{E}[V_c],
\end{align}
where 
\begin{align}\label{evcthe}
\mathbb{E}[V_c]=\mathbb{E}[T_{q_c}] + 
    \overline{\eta}_{0,c} +\frac{\varepsilon_c^{p}\,\mathbb{E}[T_{p_c}]}{1-\varepsilon_c^{p}},
   \end{align}
with 
\begin{align}\label{etatheo}
    \overline{\eta}_{0,c} = \frac{1}{\lambda_c}\!\left(1 + \sum_{j \neq c} \lambda_j\, \mathbb{E}[\gamma_{j,C+1}]\right),
\end{align}
where 
\begin{align}
\mathbb{E}[\gamma_{j,C+1}] = \frac{\varepsilon_j^{p}\,\mathbb{E}[T_{p_j}]}{1-\varepsilon_j^{p}} + \mathbb{E}[T_{q_j}],
\end{align}
where $\mathbb{E}[T_{p_j}]$ and $\mathbb{E}[T_{q_j}]$ are calculated using Lemma~\ref{lemma:condpdf}, and $\varepsilon_j^{p}$ is given in \eqref{prob_preempt}.
\end{theorem}

\begin{theorem}[Average AoI]
\label{thm:aoi}
Under the same setting as Theorem~\ref{thm:paoi}, the average AoI of source $c$ is given by
\begin{align}
    \mathbb{E}[\Delta_c] = \mathbb{E}[T_{q_c}] + \frac{\mathbb{E}[V_c^2]}{2\,\mathbb{E}[V_c]},
\end{align}
where  $\mathbb{E}[V_c]$ is given in \eqref{evcthe}, and  $\mathbb{E}[V_c^2] = \mathrm{Var}(V_c) + \mathbb{E}[V_c]^2$, with
\begin{align}
    \mathrm{Var}(V_c) = \mathrm{Var}(V_{w_c}) + \mathrm{Var}(V_{p_c}) + \mathrm{Var}(V_{q_c}),
\end{align}
and
\begin{align}
    \mathrm{Var}(V_{w_c}) &= \mathbb{E}[\eta_{0,c}^2] - \overline{\eta}_{0,c}^{\,2}, \\
    \mathrm{Var}(V_{p_c}) &= \frac{\varepsilon_c^{p}\,\mathbb{E}[T_{p_c}^2]}{1-\varepsilon_c^{p}} + \frac{(\varepsilon_c^{p})^2\,\mathbb{E}[T_{p_c}]^2}{(1-\varepsilon_c^{p})^2}, \\
    \mathrm{Var}(V_{q_c}) &= \mathbb{E}[T_{q_c}^2] - \mathbb{E}[T_{q_c}]^2,
\end{align}
where $\overline{\eta}_{0,c}$ is given in \eqref{etatheo},
$\mathbb{E}[\eta_{0,c}^2]$ is given in \eqref{eta0csecond},
$\mathbb{E}[T^2_{p_c}]$ and $\mathbb{E}[T^2_{q_c}]$ are
calculated using Lemma~\ref{lemma:condpdf}, and
$\varepsilon_c^{p}$ is given in \eqref{prob_preempt}.
\end{theorem}

Next, we  prove Theorems~\ref{thm:paoi} and \ref{thm:aoi}.
 
\begin{remark}
In the special case of a single source ($C = 1$),  the sum in \eqref{etatheo} vanishes, and the results reduce to the single-source result derived in \cite{Squ_Most_026}.
\end{remark}
 
\begin{remark}
The proposed analytical method allows each source $c$ to
have a different service-time distribution and a different preemption function $p_c(x)\in[0,1]$, provided that the 
 moments required in Theorems~\ref{thm:paoi} and~\ref{thm:aoi} are finite.
\end{remark}

\section{Derivation of the Average AoI and PAoI}
\label{analytical_sec}

The analytical method proposed for the multi-source update system consists of two steps, which we now describe in detail.
In the first step, we study the lifetime of a single update from a source $c$ that receives service, using the theory of non-homogeneous absorbing CTMCs,  as in the single-source setting \cite{Squ_Most_026}.
Using the lifetime statistics obtained in the first step, we then derive the average AoI and PAoI for source $c$ in the second step.

\subsection{Step 1: Lifetime of an Update from Source $c$}

Let $u$ be an update from source $c\in\mathcal{C}$ that currently receives service.
The \emph{lifetime} of update $u$, denoted by $T$, is defined as the total duration of time that $u$ receives service.
This service will end either because $u$ is preempted by a new arrival from source $c$, or because $u$'s service completes on its own.
Note that arrivals from sources $j \neq c$ have no effect on $u$'s service in the self-preemptive model, so the lifetime $T$ depends only on the 3-tuple $(\lambda_c, q_c(x), p_c(x))$.

\begin{figure}
\centering
\includegraphics[width=0.55\linewidth,trim = 30mm 65mm 30mm 0mm,clip]{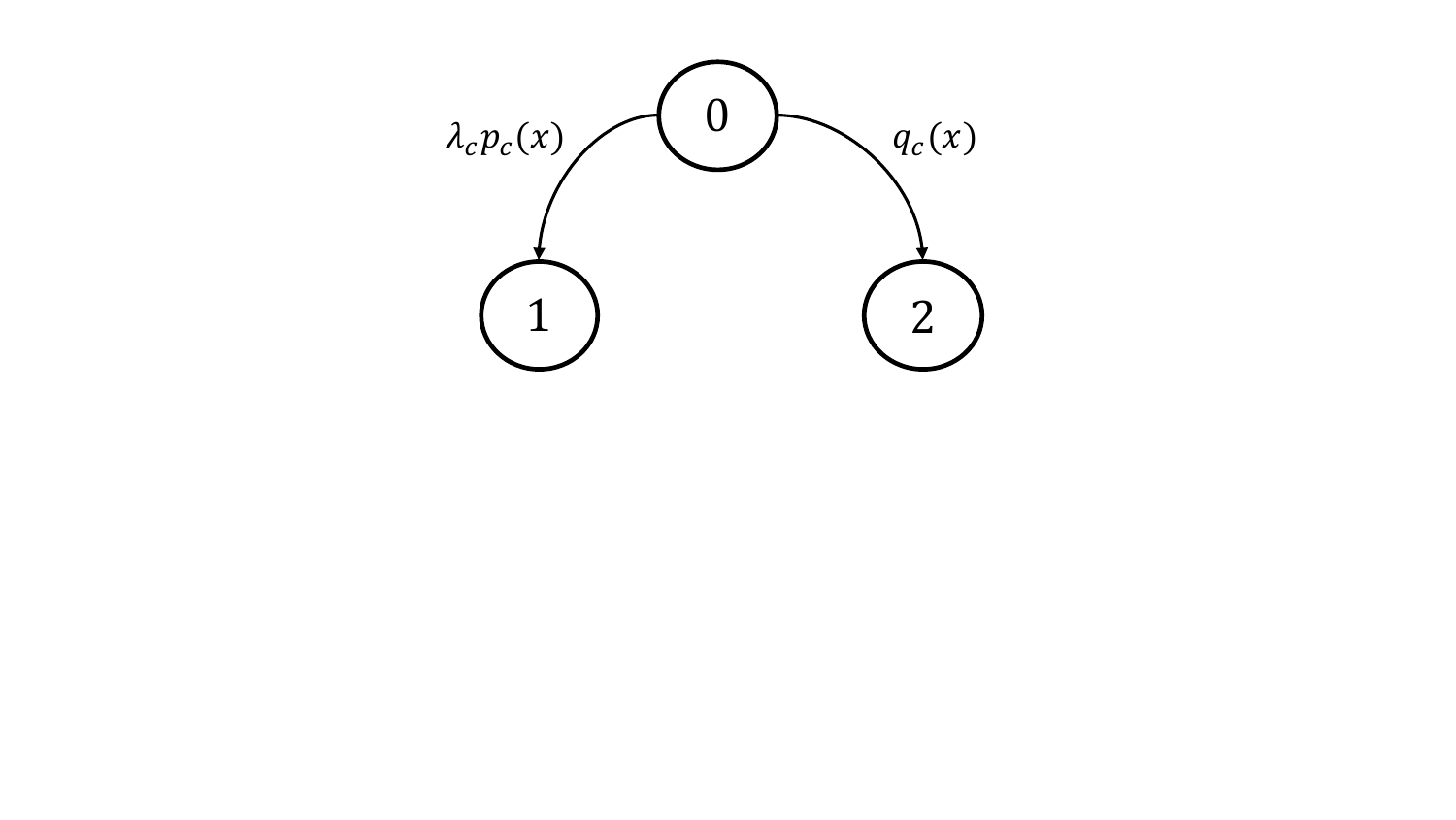}
\vspace{-5mm}
\caption{The non-homogeneous absorbing CTMC corresponding to
the lifetime $T$ of an update from source $c$.}  
\label{Ab_M_C}	
\vspace{-2mm}
\end{figure}

To study $T$, we construct a non-homogeneous absorbing CTMC
$Y(x)$ with one transient state $0$ and two absorbing states: the preemption state $1$ and the completion state $2$, as shown in Fig.~\ref{Ab_M_C}.
The CTMC $Y(x)$ starts in state $0$ at time $x = 0$ when update $u$ begins service.
It remains in state $0$ as long as neither preemption nor completion has occurred.
Absorption into state $1$ occurs when a new arrival from source $c$ preempts $u$; absorption into state $2$ occurs when $u$'s service completes without preemption.
Therefore, the lifetime $T$ equals the absorption time of $Y(x)$.

The transition rates of $Y(x)$ depend on the elapsed service time $x$, since both the preemption probability $p_c(x)$ and the service-completion hazard rate $q_c(x)$ may vary with $x$.
Hence, $Y(x)$ is a non-homogeneous CTMC. Conditioned on its current state at elapsed time $x$, its future evolution depends
only on the current state and the deterministic time parameter $x$. The non-homogeneous generator for $Y(x)$ is
\begin{align}
G(x) & = \begin{pmatrix}
-\left( q_c(x) + \lambda_c p_c(x) \right) & \lambda_c p_c(x) & q_c(x) \\
0 & 0 & 0 \\
0& 0 & 0
\end{pmatrix},
\end{align}
where rows and columns are indexed by states $0$, $1$, $2$.
The off-diagonal entries in the first row represent the rates at which $Y(x)$ leaves state $0$: the rate $\lambda_c p_c(x)$ is the rate of preemption events (a Poisson arrival from source $c$ at rate $\lambda_c$, accepted with probability $p_c(x)$), and the rate $q_c(x)$ is the instantaneous service completion rate given elapsed service time $x$.
The magnitude of the diagonal entry,
$q_c(x)+\lambda_c p_c(x)$, is the total departure rate from the transient state $0$.

The probability that $Y(x)$ remains in the transient state $0$ at time $x$, denoted by $\pi_c(x)$, satisfies the Kolmogorov forward equation and is given by
\begin{align}\label{pic}
     \pi_c(x) & =  \Pr (T > x) \\&\nonumber = {\rm e}^{- \int_0^x \left( q_c(t) + \lambda_c p_c(t) \right)  \dd{t}}, 
\end{align}
which is the unique solution to the first-order ordinary differential equation
\begin{align}
    \dv{x} \pi_c(x) &= - \left( q_c(x) + \lambda_c p_c(x) \right)\pi_c(x), \quad \pi_c(0) = 1.
\end{align}
The probability that $Y(x)$ is absorbed into the preemption state $1$ and the completion state $2$, denoted by $\varepsilon_c^{p}$ and $\varepsilon_c^{q}$, respectively, are obtained by integrating the rate of absorption over all possible absorption times:
\begin{align}
    \varepsilon_c^{p} & = \Pr(Y(\infty)=1) = \int_0^{\infty} \pi_c(x) \lambda_c p_c(x) \dd{x}, \label{prob_preempt}\\
    \varepsilon_c^{q} & = \Pr(Y(\infty)=2) = \int_0^{\infty} \pi_c(x) q_c(x) \dd{x}, \label{prob_complete}
\end{align}
and we have $\varepsilon_c^{p} + \varepsilon_c^{q} = 1$.
The following lemma derives the conditional pdf of the lifetime $T$ given each absorption state.

\begin{lemma}
\label{lemma:condpdf}
Let $T_{p_c}$ and $T_{q_c}$ denote the lifetime of update $u$ from source $c$ conditioned on absorption into the preemption state $1$ and into the completion state $2$, respectively.
For absorption events having positive probability, their pdfs are given by
\begin{align}
    f_{T_{p_c}}(x) & = \frac{\pi_c(x)\, \lambda_c p_c(x)}{\varepsilon_c^{p}}, \label{fTpc}\\
    f_{T_{q_c}}(x) & = \frac{\pi_c(x)\, q_c(x)}{\varepsilon_c^{q}}, \label{fTqc}
\end{align}
where $\pi_c(x)$ is given in \eqref{pic}.
\end{lemma}

\begin{Proof}{}
See Appendix \ref{proof-lemma:condpdf}.
\end{Proof}

When $\varepsilon_c^p=0$, the preemption event has zero probability and $T_{p_c}$ need not be defined. In this case, we have $V_{p_c}=0$, and all terms of the form $\varepsilon_c^p\mathbb{E}[T_{p_c}^m]$, $m\in\{1,2\}$, are interpreted as zero. 
For absorption events having positive probability, the first two moments of the corresponding conditional lifetimes are obtained from \eqref{fTpc} and \eqref{fTqc}
by integration. These moments are the key quantities needed for
the second step.

\subsection{Step 2: Derivation of Average AoI and PAoI}

Given the lifetime statistics from Step~1, we now derive the average AoI and PAoI for source $c$.
Recall from \eqref{meanPAoIexp} and \eqref{meanAoIexp_O} that the average AoI and PAoI depend on the first moment of $U_c$ (the system time of a successfully delivered update from source $c$) and the first two moments of $V_c$ (the inter-delivery time between consecutive successful deliveries from source $c$).
We now show how to obtain these moments.

\subsubsection{System Time $U_c$}

The system time $U_c$ of a delivered update from source $c$ equals the total time that the update spent in service, ending in a completion without preemption; therefore, $U_c \sim T_{q_c}$.
Hence,
\begin{align}
     \mathbb{E}[U_c] = \mathbb{E}[T_{q_c}]. \label{Uc}
\end{align}

\subsubsection{Decomposition of the Inter-Delivery Time $V_c$}

The inter-delivery time $V_c$ is the time elapsed between two consecutive successful deliveries from source $c$, i.e., $V_c = R_{c, k+1} - R_{c,k}$.
During $V_c$, the following events take place in order: first, the system waits until the first new update from source $c$ arrives and begins service; then, zero or more self-preemption events may occur, each consuming a preempted lifetime; finally, one update from source $c$ successfully completes service.
Accordingly, $V_c$ is decomposed as
\begin{align}
    V_c & = V_{w_c} + V_{p_c} + V_{q_c}, \label{Vdecomp}
\end{align}
where:
\begin{itemize}
    \item $V_{w_c}$: the waiting time from the last successful delivery of source $c$ until the first new update from source $c$ enters service. During this period, the server may be idle or serving updates from other sources $j \neq c$.
    \item $V_{p_c}$: the total duration of all updates from source $c$ that end up being preempted before the eventual successful delivery. This is a random sum of preempted lifetimes.
    \item $V_{q_c}$: the service time of the eventually successful update from source $c$, i.e., $V_{q_c} \sim T_{q_c}$.
\end{itemize}
The three components $V_{w_c}$, $V_{p_c}$, and $V_{q_c}$ are mutually independent, since they span disjoint intervals governed by memoryless Poisson arrivals and independent service times. Next, we characterize each component. 
\subsubsection{Characterization of $V_{q_c}$ and $V_{p_c}$}
The moments of $V_{q_c}$ follow directly from the lifetime analysis, which are given as
\begin{align}
    &\mathbb{E}[V_{q_c}]  = \mathbb{E}[T_{q_c}], ~~~~ \mathrm{Var}(V_{q_c}) = \mathbb{E}[T_{q_c}^2] - \mathbb{E}[T_{q_c}]^2. \label{Vqc}
\end{align}
The random variable $V_{p_c}$ is a random sum: it equals $\sum_{i=1}^K T_{p_c,i}$, where $K$ is the (random) number of self-preempted updates before the successful one, and $T_{p_c,i} \sim T_{p_c}$ are i.i.d.\ preempted lifetimes.
After each preemption, a fresh update from source $c$ immediately enters service, so $K$ follows a geometric distribution: with probability $\varepsilon_c^{q}$ an update in service succeeds, and with probability $\varepsilon_c^{p}$ it is preempted. Thus,
\begin{align}
    \Pr(K=k) & = \varepsilon_c^{q}(\varepsilon_c^{p})^k, \qquad k=0,1,\ldots .
\end{align}
Consequently, the first moment and the second factorial moment of $K$ are
\begin{align}
    &\mathbb{E}[K]  = \frac{\varepsilon_c^{p}}{1-\varepsilon_c^{p}},
    \\&
    \mathbb{E}[K(K-1)] = \frac{2(\varepsilon_c^{p})^2}{(1-\varepsilon_c^{p})^2}.
\end{align}
Using the standard moment formulas for a random sum of i.i.d.\ random variables, with $K$ independent of the preempted lifetimes, we obtain
\begin{align}\label{EVpc}
    \mathbb{E}[V_{p_c}]
    & = \mathbb{E}[K]\mathbb{E}[T_{p_c}]
\\&\nonumber=\frac{\varepsilon_c^{p} \mathbb{E}[T_{p_c}]}{1-\varepsilon_c^{p}}, 
\end{align}
and
\begin{align}\label{EVpc2}
    \mathbb{E}[V_{p_c}^2]
    & = \mathbb{E}[K]\mathbb{E}[T_{p_c}^2]
    + \mathbb{E}[K(K-1)]\mathbb{E}[T_{p_c}]^2
    \\
    &\nonumber =\frac{2 (\varepsilon_c^{p})^2 \mathbb{E}[T_{p_c}]^2}{(1-\varepsilon_c^{p})^2}
    + \frac{\varepsilon_c^{p} \mathbb{E}[T_{p_c}^2]}{1-\varepsilon_c^{p}}. 
\end{align}

Finally, $\mathrm{Var}(V_{p_c})$ is given as
\begin{align}\label{var_vpc}
    \mathrm{Var}(V_{p_c}) &= \frac{\varepsilon_c^{p}\,\mathbb{E}[T_{p_c}^2]}{1-\varepsilon_c^{p}} + \frac{(\varepsilon_c^{p})^2\,\mathbb{E}[T_{p_c}]^2}{(1-\varepsilon_c^{p})^2}.
\end{align}

\subsubsection{Characterization of $V_{w_c}$}
 
The waiting time $V_{w_c}$ is the most involved component in the multi-source setting.
After source $c$'s successful delivery, the server becomes free.
The next update from source $c$ can enter service only once it arrives \emph{and} the server is available.
However, between this moment and the moment when source $c$'s next update enters service, the server may be occupied serving one or more updates from other sources $j \neq c$, e.g., upon completing service for some source $j \neq c$, the server may again be occupied by yet another source $j' \neq c$ before source $c$ arrives, and this can repeat an arbitrary number of times.
This queueing dynamic needs to be tracked carefully.
 
\begin{figure}
\centering
\includegraphics[width=.65\linewidth,trim = 0mm 0mm 0mm 0mm,clip]{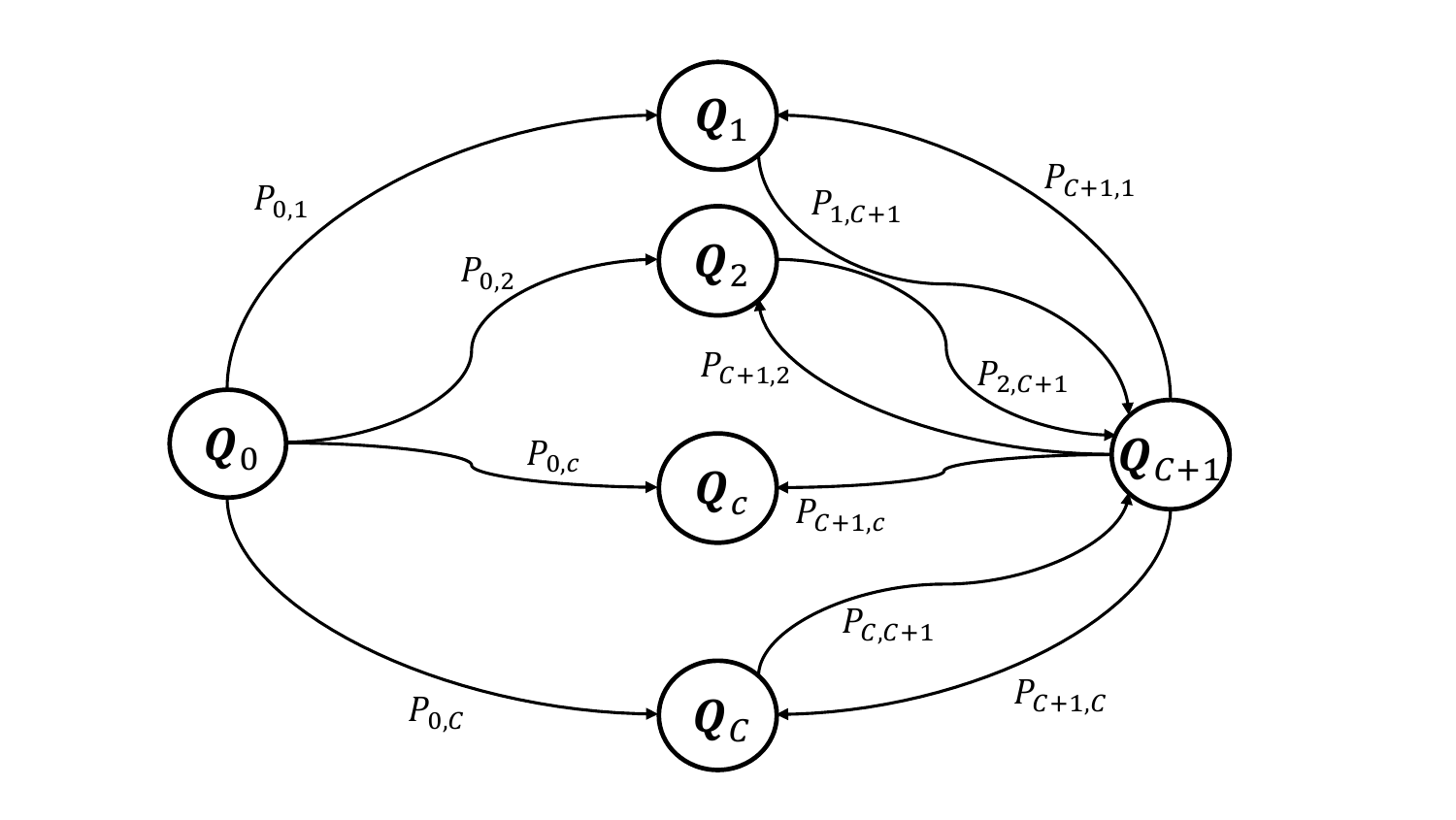}
\vspace{-5mm}
\caption{The semi-Markov chain corresponding to the waiting time until a new packet from source $c$ enters service.}  
\label{semi-Markov chain}	
\vspace{-2mm}
\end{figure}

To characterize $V_{w_c}$, we construct a semi-Markov chain with state space
\begin{align}
    \mathcal{S} & = \{ \boldsymbol{Q}_0, \boldsymbol{Q}_1, \ldots, \boldsymbol{Q}_C, \boldsymbol{Q}_{C+1} \}.
\end{align}
The semi-Markov chain is shown in Fig.~\ref{semi-Markov chain}. The states are defined as follows:
\begin{itemize}
    \item $\boldsymbol{Q}_0$: a packet from source $c$ (the source of interest) has just been successfully delivered, and the server is now idle.
    This is the \emph{initial} state of the chain at the beginning of each $V_{w_c}$ interval.
    \item $\boldsymbol{Q}_j$ ($j = 1, \ldots, C$, $j \neq c$): a packet from source $j$ is currently in service.
    \item $\boldsymbol{Q}_c$: a packet from source $c$ has just entered service.
    This is the \emph{target} state; the first passage to $\boldsymbol{Q}_c$ marks the end of $V_{w_c}$.
    \item $\boldsymbol{Q}_{C+1}$: a packet from a source other than $c$ has just been delivered, and the server is again idle.
    This state has the same structure as $\boldsymbol{Q}_0$ from the server's perspective, but is reached via a delivery from source $j \neq c$ rather than from source $c$.
\end{itemize}

The chain starts in $\boldsymbol{Q}_0$ immediately after source $c$'s delivery, and we seek the first passage time $\eta_{0,c}$ from $\boldsymbol{Q}_0$ to the target state $\boldsymbol{Q}_c$, which equals $V_{w_c}$.
Note that $\boldsymbol{Q}_0$ and $\boldsymbol{Q}_{C+1}$ are structurally identical idle states: in both, the server is free and waiting for the next arrival.
Their distinction is only used to track where the chain was before becoming idle, and both follow the same transition dynamics.
Accordingly, the transition probabilities and sojourn times out of $\boldsymbol{Q}_0$ and $\boldsymbol{Q}_{C+1}$ are identical.

The transitions and sojourn times of the semi-Markov chain are characterized as follows.
 
\textbf{From $\boldsymbol{Q}_0$ and $\boldsymbol{Q}_{C+1}$ (idle states):}
The server is free, so the next event is the arrival of the first packet from any source.
Since source arrivals are independent Poisson processes, the first arrival is from source $j$ with probability $\lambda_j / \lambda$.
If $j = c$, the chain transitions directly to the target $\boldsymbol{Q}_c$; otherwise, it transitions to $\boldsymbol{Q}_j$.
The transition probabilities are therefore
\begin{align}
    P_{0,j} = P_{C+1,j} & = \frac{\lambda_j}{\lambda}, \quad j = 1, \ldots, C. \label{P0j}
\end{align}
The following lemma characterizes the sojourn time distribution in the idle states.
 
\begin{lemma}
\label{lemma:sojourn}
The pdf of the sojourn time $\gamma_{0,j}$ (resp.\ $\gamma_{C+1,j}$) spent in the idle state $\boldsymbol{Q}_0$ (resp.\ $\boldsymbol{Q}_{C+1}$) before transitioning to state $\boldsymbol{Q}_j$, for any $j = 1,\ldots,C$, is exponential with rate $\lambda$, which is given as
\begin{align}
    f_{\gamma_{0,j}}(t) = f_{\gamma_{C+1,j}}(t) & = \lambda\, {\rm e}^{-\lambda t}, \quad t \geq 0. \label{gamma0j}
\end{align}
Consequently, the first two moments are given as
\begin{align}
    &\mathbb{E}[\gamma_{0,j}] = \mathbb{E}[\gamma_{C+1,j}] = \frac{1}{\lambda}, \\&
    \mathbb{E}[\gamma_{0,j}^2] = \mathbb{E}[\gamma_{C+1,j}^2] = \frac{2}{\lambda^2}. \label{gamma0j_moments}
\end{align}
\end{lemma}
 \begin{Proof}
 Since the source arrival processes are independent Poisson
processes, their superposition is a Poisson process with rate
$\lambda=\sum_{k=1}^C\lambda_k$. Hence, the time to the next
arrival is exponentially distributed with rate $\lambda$.
Moreover, the minimum arrival time is independent of the identity
of the source generating the first arrival. Therefore, conditioning
on source $j$ being the first to arrive does not change this
distribution, which gives \eqref{gamma0j}.
 \end{Proof}

\textbf{From $\boldsymbol{Q}_j$ ($j \neq c$):}
A packet from source $j$ is in service. While source $j$ occupies the server, arrivals from source $c$ are discarded. A source-$c$ update can enter service only after source $j$'s service episode ends.
Source $j$'s update may be preempted by an arrival from source $j$, or it may complete on its own; either way, the total duration of source $j$'s consecutive occupancy of the server, counting all self-preemptions from source $j$, up to the eventual successful completion, equals $\gamma_{j,C+1} = V_{p_j} + V_{q_j}$, where $V_{p_j}$ is the total preempted duration and $V_{q_j}$ is the final successful service time.
After source $j$'s episode ends by a successful delivery, the server becomes idle again, transitioning to $\boldsymbol{Q}_{C+1}$.
Hence, the transition from $\boldsymbol{Q}_j$ always goes to $\boldsymbol{Q}_{C+1}$:
\begin{align}
    P_{j,C+1} & = 1, \quad j = 1, \ldots, C,\; j \neq c, \label{PjC1}
\end{align}
with sojourn time equal to the full service episode of source $j$. Using the lifetime analysis, similar results as in  \eqref{Vqc}, \eqref{EVpc}, and \eqref{EVpc2} are concluded for $V_{p_j}$ and $V_{q_j}$, and thus, we have 
\begin{align}
    \mathbb{E}[\gamma_{j,C+1}] & = \mathbb{E}[V_{p_j}] + \mathbb{E}[V_{q_j}], \label{gammajC1_mean}\\
    \mathbb{E}[\gamma_{j,C+1}^2] & =  \mathbb{E}[V_{p_j} ^2]+\mathbb{E}[ V_{q_j}^2]+2\mathbb{E}[V_{p_j}]\mathbb{E}[V_{q_j}],\label{gammajC1_second}
\end{align}
where 
\begin{align}
    &\mathbb{E}[V_{q_j}]  = \mathbb{E}[T_{q_j}],
    \\&
    \mathbb{E}[V^2_{q_j}] = \mathbb{E}[T_{q_j}^2],\\&
    \mathbb{E}[V_{p_j}] 
    =\frac{\varepsilon_j^{p} \mathbb{E}[T_{p_j}]}{1-\varepsilon_j^{p}}, 
    \\&
    \mathbb{E}[V_{p_j}^2]  
    =\frac{\varepsilon_j^{p} \mathbb{E}[T_{p_j}^2]}{1-\varepsilon_j^{p}}+\frac{2 (\varepsilon_j^{p})^2 \mathbb{E}[T_{p_j}]^2}{(1-\varepsilon_j^{p})^2}.  
\end{align}

\textbf{Target state $\boldsymbol{Q}_c$:}
This state is absorbing for the purpose of computing $V_{w_c}$: the first passage time to $\boldsymbol{Q}_c$ terminates the $V_{w_c}$ interval, and a new $V_{w_c}$ interval will begin only after source $c$ eventually completes its current service episode and triggers a new $\boldsymbol{Q}_0$ entry.
Since $\boldsymbol{Q}_0$ and $\boldsymbol{Q}_{C+1}$ have identical dynamics, we denote their common first passage time to $\boldsymbol{Q}_c$ by $\eta_{0,c}$, which satisfies $\mathbb{E}[\eta_{0,c}] = \mathbb{E}[\eta_{C+1,c}]$ and $\mathbb{E}[\eta_{0,c}^2] = \mathbb{E}[\eta_{C+1,c}^2]$.
 
\subsubsection{First and Second Moments of the First Passage Time $\eta_{0,c}$}
 
Let $\eta_{i,j}$ denote the first passage time from state $\boldsymbol{Q}_i$ to state $\boldsymbol{Q}_j$ in the semi-Markov chain, and let $\overline{\eta}_{i,j} = \mathbb{E}[\eta_{i,j}]$.
By standard semi-Markov chain theory \cite{kleinrock_book}, the first moments of the first passage times satisfy the following system of linear equations
\begin{align}
    \overline{\eta}_{i,j} & = \sum_{k} P_{i,k} \left( \mathbb{E}[\gamma_{i,k}] + \overline{\eta}_{k,j} \right),  ~~ \text{for}~~ i \neq j,\label{fptfirst}
\end{align}
with $\eta_{j,j}=0$ and hence $\overline{\eta}_{j,j} = 0$ for all $j$.
Similarly, the second moments satisfy
\begin{align}
    \mathbb{E}[\eta_{i,j}^2] & = \sum_{k} P_{i,k}\, \mathbb{E}\!\left[ \left(\gamma_{i,k} + \eta_{k,j}\right)^2 \right], \quad i \neq j. \label{fptsecond}
\end{align}
We now apply \eqref{fptfirst} and \eqref{fptsecond} to the semi-Markov chain constructed above.
The following lemma provides closed-form expressions for $\overline{\eta}_{0,c}$ and $\mathbb{E}[\eta_{0,c}^2]$.
 
\begin{lemma}
\label{lemma:fpt}
For the semi-Markov chain with state space $\{\boldsymbol{Q}_0, \boldsymbol{Q}_1, \ldots, \boldsymbol{Q}_C, \boldsymbol{Q}_{C+1}\}$ and transition structure defined in \eqref{P0j}--\eqref{gammajC1_second}
the first and second moments of the first passage time $\eta_{0,c}$ from the idle state $\boldsymbol{Q}_0$ to the target state $\boldsymbol{Q}_c$ are given by
\begin{align}
    \overline{\eta}_{0,c}
    & = \frac{1}{\lambda_c}\!\left(1 + \sum_{j \neq c} \lambda_j \,\mathbb{E}[\gamma_{j,C+1}]\right), \label{eta0c}
\end{align}
\begin{align}
    \mathbb{E}[\eta_{0,c}^2]
    & = \frac{2}{\lambda \lambda_c} + \sum_{j \neq c} \frac{\lambda_j}{\lambda_c} \mathbb{E}[\gamma_{j,C+1}^2]
    + 2\!\left(\frac{1}{\lambda} + \overline{\eta}_{0,c}\right) 
    \sum_{j \neq c} \frac{\lambda_j}{\lambda_c} \mathbb{E}[\gamma_{j,C+1}]  + 2\!\left(\frac{\lambda}{\lambda_c} - 1\right)\frac{1}{\lambda}\, \overline{\eta}_{0,c}, \label{eta0csecond}
\end{align}
where $\mathbb{E}[\gamma_{j,C+1}]$ and $\mathbb{E}[\gamma^2_{j,C+1}]$ are given in \eqref{gammajC1_mean} and \eqref{gammajC1_second}, respectively.
\end{lemma}
 
\begin{Proof}{}
\textbf{First moment.}
We apply \eqref{fptfirst} from state $\boldsymbol{Q}_0$ with target $\boldsymbol{Q}_c$.
From $\boldsymbol{Q}_0$, the chain transitions to $\boldsymbol{Q}_c$ directly, if source $c$ arrives first (with probability $\lambda_c/\lambda$) or to $\boldsymbol{Q}_j$ for some $j \neq c$. Thus, we have
\begin{align}
    \overline{\eta}_{0,c} & = P_{0,c}\,\mathbb{E}[\gamma_{0,c}] + \sum_{j \neq c} P_{0,j}\!\left(\mathbb{E}[\gamma_{0,j}] + \overline{\eta}_{j,c}\right). \label{proof_eta0c_raw}
\end{align}
From $\boldsymbol{Q}_j$ ($j \neq c$), the chain always transitions to $\boldsymbol{Q}_{C+1}$ after source $j$'s service episode, and since $\overline{\eta}_{C+1,c} = \overline{\eta}_{0,c}$ by symmetry, the first passage time from $\boldsymbol{Q}_j$ decomposes as
\begin{align}
    \overline{\eta}_{j,c} & = \mathbb{E}[\gamma_{j,C+1}] + \overline{\eta}_{0,c}. \label{proof_etajc}
\end{align}
Substituting \eqref{proof_etajc} into \eqref{proof_eta0c_raw}, and using $P_{0,c} = \lambda_c/\lambda$, $P_{0,j} = \lambda_j/\lambda$, and $\mathbb{E}[\gamma_{0,j}] = 1/\lambda$ for all $j$, we obtain
\begin{align}
    \overline{\eta}_{0,c} & = \frac{\lambda_c}{\lambda}\cdot\frac{1}{\lambda} + \sum_{j \neq c} \frac{\lambda_j}{\lambda}\!\left(\frac{1}{\lambda} + \mathbb{E}[\gamma_{j,C+1}] + \overline{\eta}_{0,c}\right) \nonumber \\
    & = \frac{1}{\lambda} + \sum_{j \neq c}\frac{\lambda_j}{\lambda}\mathbb{E}[\gamma_{j,C+1}] + \left(1 - \frac{\lambda_c}{\lambda}\right)\overline{\eta}_{0,c}. \label{proof_eta0c_intermediate}
\end{align}
Collecting the $\overline{\eta}_{0,c}$ terms on the left-hand side and dividing both sides by $\lambda_c / \lambda$ yields \eqref{eta0c}.
 
\textbf{Second moment.}
We apply \eqref{fptsecond} from state $\boldsymbol{Q}_0$ with target $\boldsymbol{Q}_c$:
\begin{align}
    \mathbb{E}[\eta_{0,c}^2] &= P_{0,c}\,\mathbb{E}[\gamma_{0,c}^2]
    + \sum_{j \neq c} P_{0,j}\,\mathbb{E}\!\left[(\gamma_{0,j} + \eta_{j,c})^2\right]. \label{proof_sec_raw}
\end{align}
Since $\gamma_{0,c}$ is exponential with rate $\lambda$, we have $\mathbb{E}[\gamma_{0,c}^2] = 2/\lambda^2$.
For the terms $j \neq c$, we expand the square and use the independence of $\gamma_{0,j}$ and $\eta_{j,c}$:
\begin{align}
    \mathbb{E}\!\left[(\gamma_{0,j} + \eta_{j,c})^2\right]
    &= \mathbb{E}[\gamma_{0,j}^2] + 2\,\mathbb{E}[\gamma_{0,j}]\,\mathbb{E}[\eta_{j,c}] + \mathbb{E}[\eta_{j,c}^2]. \label{proof_expand_sq}
\end{align}
From state $\boldsymbol{Q}_j$ ($j \neq c$), the chain always returns to an idle state after a sojourn of $\gamma_{j,C+1}$, so the first passage time from $\boldsymbol{Q}_j$ to the target satisfies $\eta_{j,c} \sim (\gamma_{j,C+1} + \eta_{0,c})$ (with $\gamma_{j,C+1}$ and $\eta_{0,c}$ independent). Therefore,
\begin{align}
    \mathbb{E}[\eta_{j,c}] &= \mathbb{E}[\gamma_{j,C+1}] + \overline{\eta}_{0,c}, \label{proof_etajc_first}\\
    \mathbb{E}[\eta_{j,c}^2] &= \mathbb{E}[\gamma_{j,C+1}^2] + 2\,\mathbb{E}[\gamma_{j,C+1}]\,\overline{\eta}_{0,c} + \mathbb{E}[\eta_{0,c}^2]. \label{proof_etajc_second}
\end{align}
Substituting \eqref{proof_etajc_first} and \eqref{proof_etajc_second} into \eqref{proof_expand_sq}, and using $\mathbb{E}[\gamma_{0,j}] = 1/\lambda$ and $\mathbb{E}[\gamma_{0,j}^2] = 2/\lambda^2$, we obtain
\begin{align}\nonumber
    \mathbb{E}\!\left[(\gamma_{0,j} + \eta_{j,c})^2\right]
    &= \frac{2}{\lambda^2} + \frac{2}{\lambda}\!\left(\mathbb{E}[\gamma_{j,C+1}] + \overline{\eta}_{0,c}\right)
    \\& +\mathbb{E}[\gamma_{j,C+1}^2] + 2\,\mathbb{E}[\gamma_{j,C+1}]\,\overline{\eta}_{0,c} + \mathbb{E}[\eta_{0,c}^2]. \label{proof_expand_full}
\end{align}
Substituting \eqref{proof_expand_full} into \eqref{proof_sec_raw} with $P_{0,c} = \lambda_c/\lambda$ and $P_{0,j} = \lambda_j/\lambda$, collecting the $\mathbb{E}[\eta_{0,c}^2]$ terms, dividing both sides by $\lambda_c/\lambda$, and simplifying using $\sum_{j\neq c}\lambda_j/\lambda = 1 - \lambda_c/\lambda$ yields \eqref{eta0csecond}.
\end{Proof}
 
Since $V_{w_c}$ is the first passage time $\eta_{0,c}$ from $\boldsymbol{Q}_0$ to $\boldsymbol{Q}_c$ in the semi-Markov chain, we have
\begin{align}\label{Vw_moments}
    &\mathbb{E}[V_{w_c}]  = \overline{\eta}_{0,c}, ~~~~~
     \mathrm{Var}(V_{w_c})= \mathbb{E}[\eta_{0,c}^2] -\overline{\eta}
    ^2_{0,c}. 
\end{align}
 
\subsubsection{Final Computation of Average AoI and PAoI}
 
Having obtained the moments of all three components of $V_c$ from \eqref{Vqc}, \eqref{EVpc}, \eqref{var_vpc}, and \eqref{Vw_moments}, and using their mutual independence, we compute the moments of $V_c = V_{w_c} + V_{p_c} + V_{q_c}$ as follows:
\begin{align}
    \mathbb{E}[V_c] & = \mathbb{E}[V_{w_c}] + \mathbb{E}[V_{p_c}] + \mathbb{E}[V_{q_c}], \label{meanVc}\\
    \mathrm{Var}(V_c) & = \mathrm{Var}(V_{w_c}) + \mathrm{Var}(V_{p_c}) + \mathrm{Var}(V_{q_c}), \label{varVc}\\
    \mathbb{E}[V_c^2] & = \mathrm{Var}(V_c) + \mathbb{E}[V_c]^2. \label{secondmomentVc}
\end{align}
Finally, substituting $\mathbb{E}[U_c] = \mathbb{E}[T_{q_c}]$ from \eqref{Uc}, together with $\mathbb{E}[V_c]$ from \eqref{meanVc} and $\mathbb{E}[V_c^2]$ from \eqref{secondmomentVc}, into the AoI and PAoI formulas \eqref{meanAoIexp_O} and \eqref{meanPAoIexp}, we obtain the average AoI $\mathbb{E}[\Delta_c]$ and average PAoI $\mathbb{E}[\Phi_c]$ for source $c$, establishing Theorems~\ref{thm:paoi} and \ref{thm:aoi}.

\section{Numerical Results for the Single-Source Case}\label{results_sec}
In this section, for the single-source case, we quantify the reductions in average AoI and PAoI obtained by the FP policy and by the tuned PP and TP policies, relative to the NP policy, for fifteen service-time distributions spanning increasing, unimodal, and decreasing hazard-rate shapes. This single-source study complements the general-source analysis by identifying when elapsed-service-time-dependent preemption is beneficial.

\subsection{Service Time Distributions}
\label{sec:distributions}
\begin{figure*}[tb]
    \centering
    \includegraphics[width=1\linewidth]{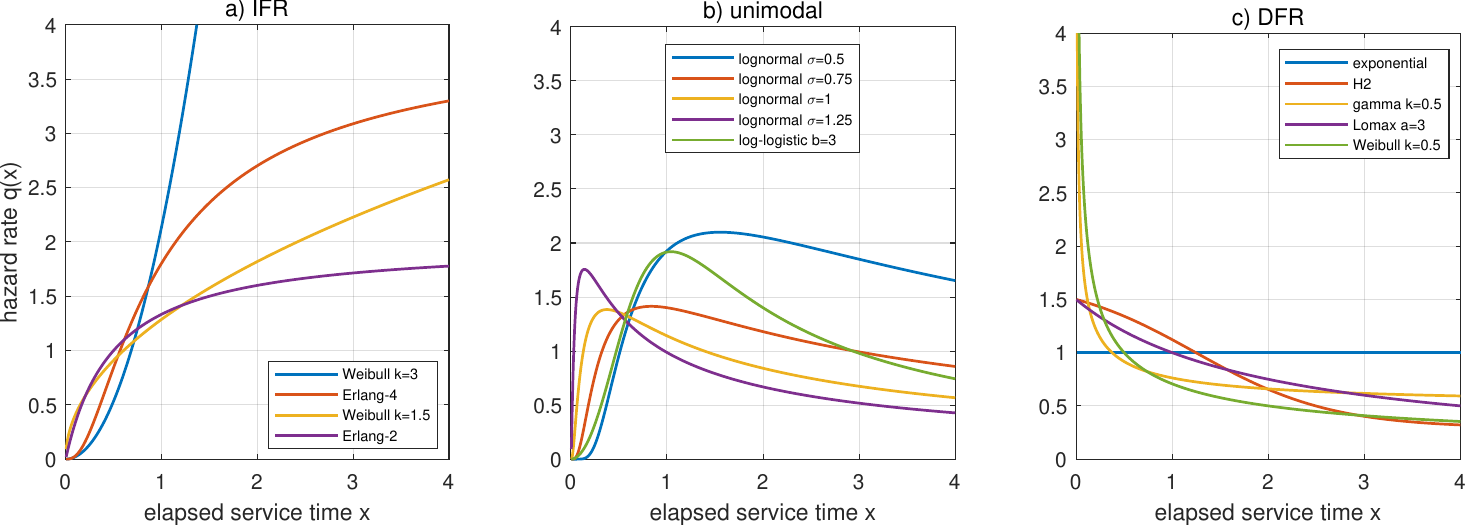}
    \caption{Hazard rates $q(x)$ of the service time distributions of Section~\ref{sec:distributions}, all with $\mathbb{E}[S]=1$: a) IFR b) unimodal c) DFR. The deterministic service time (hazard rate concentrated at $x=1$) is not drawn.}
    \label{fig:hazards}
\end{figure*}

We consider fifteen distributions, all normalized to $\mathbb{E}[S]=1$ which are grouped by the shape of the hazard rate $q(x)$ as follows (see \cite{book_oconnor} for the general forms and for the moments).
\begin{itemize}
\item \emph{Increasing failure rate (IFR)}: $q(x)$ is non-decreasing, i.e., the longer an update has been in service, the more likely it is to complete soon. We use the deterministic service time $S=1$; the Erlang-$k$ distribution with rate $k$, \begin{align} 
f(x)=\dfrac{k^k x^{k-1}{\rm e}^{-kx}}{(k-1)!},~~~k\in\{2,4\};
\end{align}
and the Weibull distribution 
\begin{align} 
f(x)=\frac{k}{c}\left(\frac{x}{c}\right)^{k-1}{\rm e}^{-(x/c)^k},
\end{align}
with hazard rate $q(x)=\frac{k}{c}\left(\frac{x}{c}\right)^{k-1}$, scale $c=1/\Gamma(1+1/k)$ and shape $k\in\{1.5,3\}$.
\item \emph{Unimodal hazard rate}: $q(0)=0$, $q(x)$ increases to a maximum and then decreases to zero, so that retaining an update during the increasing-hazard phase can be beneficial, whereas preempting an update in the decreasing-hazard phase may improve information freshness. We use the lognormal distribution
\begin{align}
f(x)=\frac{1}{x\sigma\sqrt{2\pi}}{\rm e}^{-(\ln x-\mu)^2/(2\sigma^2)}, 
\end{align}
whose mean is ${\rm e}^{\mu+\sigma^2/2}$, with $\mu=-\sigma^2/2$ and $\sigma\in\{0.5,0.75,1,1.25\}$ with squared coefficient of variation (variance divided by mean squared) $\mathrm{CV}^2={\rm e}^{\sigma^2}-1$, and the log-logistic distribution 
\begin{align}
f(x)=\frac{b}{c}\left(\frac{x}{c}\right)^{b-1}\big/\big(1+(x/c)^b\big)^2,
\end{align}
complementary cdf (ccdf) $\bar{F}(x)=1-F(x)=1/(1+(x/c)^b)$, with $c=b\sin(\pi/b)/\pi$ and $b=3$.
\item \emph{Decreasing failure rate (DFR)}: $q(x)$ is non-increasing, i.e., the longer an update has lasted, the longer it is expected to last. We use the exponential distribution; the two-phase hyper-exponential (H$_2$) distribution 
\begin{align}
f(x)=p_1\mu_1{\rm e}^{-\mu_1x}+p_2\mu_2{\rm e}^{-\mu_2x},
\end{align}
with balanced means $p_1/\mu_1=p_2/\mu_2=1/2$ and $\mathrm{CV}^2=3$, i.e., $p_1=\frac12\big(1+\sqrt{1/2}\big)$, $p_2=1-p_1$, $\mu_i=2p_i$; the gamma distribution 
\begin{align}
f(x)=\sqrt{0.5}\,x^{-1/2}{\rm e}^{-x/2}/\Gamma(0.5),
\end{align}
with shape $k=0.5$ and rate $0.5$; 
the Lomax (Pareto type II) distribution 
\begin{align}
\bar{F}(x)=(1+x/\beta)^{-a},
\end{align}
with hazard rate $q(x)=a/(\beta+x)$, $a=3$ and $\beta=a-1=2$; and the Weibull distribution above with $k=0.5$.
\end{itemize}
The hazard rates of the fifteen distributions are depicted in Fig.~\ref{fig:hazards}, and their $\mathrm{CV}^2$ values are listed in Table~\ref{tab:gains}. We recall that IFR implies $\mathrm{CV}^2\le1$ and DFR implies $\mathrm{CV}^2\ge1$ \cite{barlow65}, but not conversely. On the other hand, the unimodal-hazard distributions appear on both sides of $\mathrm{CV}^2=1$.

\subsection{Computation and Policy Tuning}
\label{sec:method}
 We evaluate the numerical integrals with MATLAB's adaptive quadrature \texttt{integral} with absolute and relative tolerances $10^{-12}$ and $10^{-9}$, respectively.
The numerically optimized parameters $\alpha^*$ (PP) and $\tau^*$ (TP) are obtained by a one-dimensional search in two stages. First, the objective is evaluated on a coarse grid of the parameter which are 48 equally spaced thresholds on $[0,8\,\mathbb{E}[S]]$ for TP, and 40 logarithmically spaced probabilities on $[10^{-4},1]$ for PP. Second, the best grid point and its two neighbors define a bracket on which MATLAB's \texttt{fminbnd} is run where \texttt{fminbnd} implements Brent's method \cite{NumericalRecipes}, i.e., golden section search combined with parabolic interpolation, and searches for a minimum within the selected bracket. We use the tolerance $10^{-6}$ on the parameter. The coarse scan makes the procedure robust in the absence of a unimodality guarantee, and costs only a few dozen additional evaluations. In the numerical cases considered, the objectives appear unimodal in $\alpha$ under PP and in $\tau$ under TP; this is an empirical observation rather than a general optimality guarantee. 

\subsection{Results}
\label{sec:results}
\begin{figure*}[t]
\vspace*{0.1cm}
    \centering
    \includegraphics[width=1\linewidth]{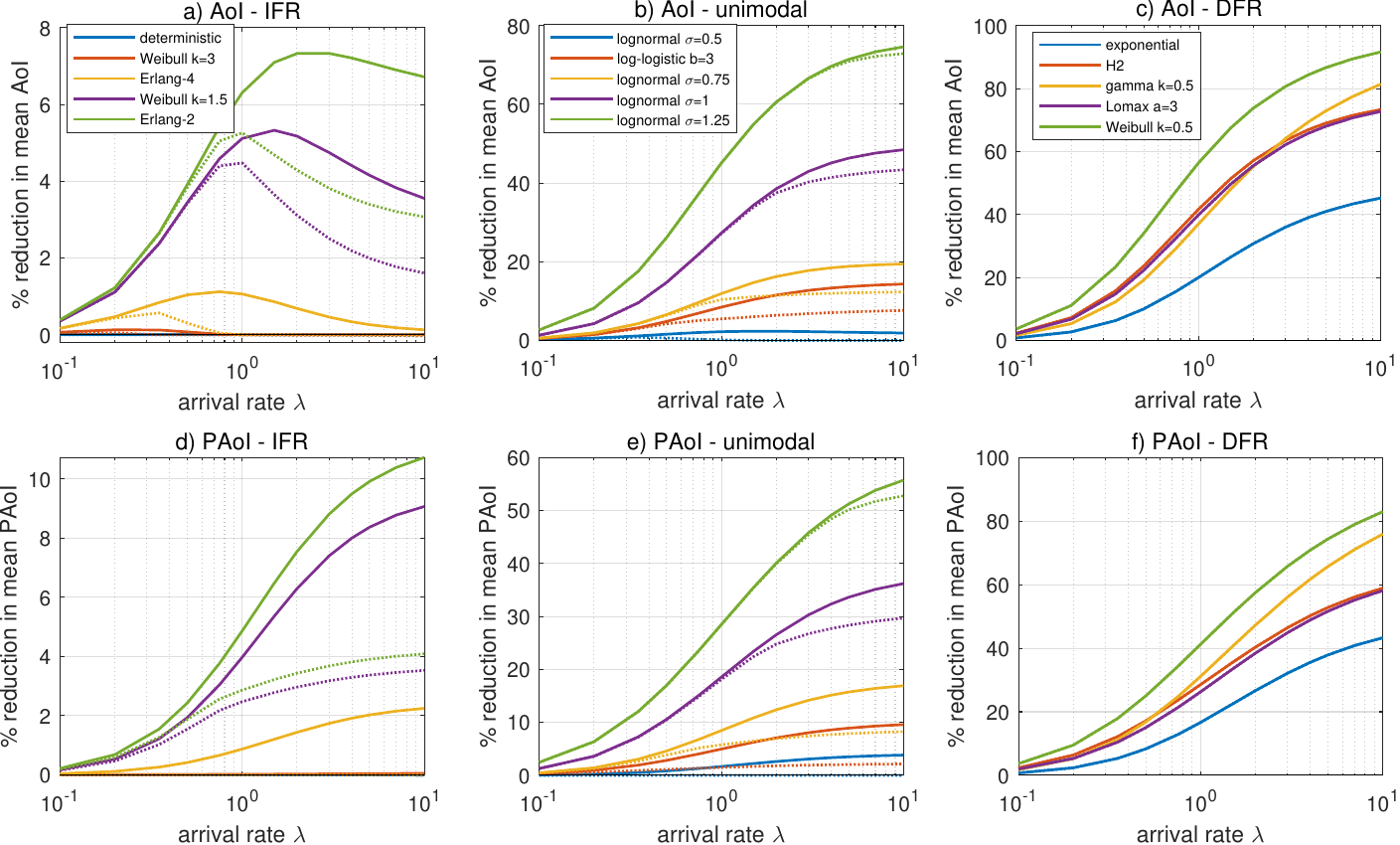}
    \caption{Percentage reduction in average AoI (top row) and average PAoI (bottom row) with respect to NP, of the tuned TP policy (solid) and the tuned PP policy (dotted), as a function of $\lambda$, for IFR, unimodal, and DFR groups when $\mathbb{E}[S]=1$.}
    \label{fig:gains}
    \vspace{-2mm}
\end{figure*}

Fig.~\ref{fig:gains} depicts the percentage reduction in average AoI (first row) and average PAoI (second row) with respect to NP, i.e., $100\,(\mathbb{E}[\cdot]_{\rm NP}-\mathbb{E}[\cdot]_{\rm policy})/\mathbb{E}[\cdot]_{\rm NP}$, for the tuned TP policy (solid lines) and the tuned PP policy (dotted lines), as a function of the arrival rate $\lambda\in[0.1,10]$, with one plot for each hazard rate group. In Table~\ref{tab:gains}, we report the same percentage reductions together with those of the FP policy, and the numerically optimized parameters $\alpha^*$ and $\tau^*$ when $\lambda$ is fixed to 5.

\begin{table*}[tb]
\centering
\caption{Percentage reduction in average AoI and PAoI with respect to NP at $\lambda=5$ ($\mathbb{E}[S]=1$) for FP, tuned PP (with $\alpha^*$) and tuned TP (with $\tau^*$). Entries $<\!-99$ indicate that FP is worse than NP by more than 99\%; -- indicates that the tuned policy coincides with NP.}
\label{tab:gains}
\setlength{\tabcolsep}{4pt}
\begin{tabular}{|l|c|rrrrr|rrrrr|}
\hline
 & & \multicolumn{5}{c|}{average AoI} & \multicolumn{5}{c|}{average PAoI} \\
service time & $\mathrm{CV}^2$ & FP & PP & $\alpha^*$ & TP & $\tau^*$ & FP & PP & $\alpha^*$ & TP & $\tau^*$ \\ \hline
deterministic & 0.00 & $<\!-99$ & 0.0 & -- & 0.0 & -- & $<\!-99$ & 0.0 & -- & 0.0 & -- \\
Weibull, $k=3$ & 0.13 & $<\!-99$ & 0.0 & -- & 0.0 & -- & $<\!-99$ & 0.0 & -- & 0.0 & -- \\
Erlang-4 & 0.25 & $<\!-99$ & 0.0 & -- & 0.3 & 1.90 & $<\!-99$ & 0.0 & -- & 2.0 & 1.40 \\
Weibull, $k=1.5$ & 0.46 & $-26.4$ & 2.0 & 0.12 & 4.2 & 1.08 & $-15.7$ & 3.4 & 0.18 & 8.4 & 0.89 \\
Erlang-2 & 0.50 & $-34.2$ & 3.4 & 0.14 & 7.1 & 0.99 & $-24.4$ & 3.9 & 0.17 & 9.9 & 0.87 \\
\hline
lognormal, $\sigma=0.5$ & 0.28 & $<\!-99$ & 0.0 & -- & 2.1 & 1.51 & $<\!-99$ & 0.0 & -- & 3.5 & 1.30 \\
log-logistic, $b=3$ & 0.65 & $<\!-99$ & 7.2 & 0.09 & 13.7 & 1.20 & $<\!-99$ & 2.0 & 0.06 & 8.9 & 1.12 \\
lognormal, $\sigma=0.75$ & 0.76 & $-35.7$ & 12.1 & 0.19 & 18.8 & 0.81 & $-34.5$ & 7.9 & 0.17 & 15.8 & 0.79 \\
lognormal, $\sigma=1$ & 1.72 & 38.1 & 42.1 & 0.51 & 46.3 & 0.35 & 23.7 & 28.3 & 0.48 & 33.6 & 0.36 \\
lognormal, $\sigma=1.25$ & 3.77 & 70.9 & 70.9 & 1 & 71.4 & 0.10 & 50.1 & 50.1 & 1 & 51.2 & 0.12 \\
\hline
exponential & 1.00 & 41.0 & 41.0 & 1 & 41.0 & 0 & 37.9 & 37.9 & 1 & 37.9 & 0 \\
H$_2$ & 3.00 & 69.0 & 69.0 & 1 & 69.0 & 0 & 52.8 & 52.8 & 1 & 52.8 & 0 \\
gamma, $k=0.5$ & 2.00 & 72.9 & 72.9 & 1 & 72.9 & 0 & 65.7 & 65.7 & 1 & 65.7 & 0 \\
Lomax, $a=3$ & 3.00 & 68.0 & 68.0 & 1 & 68.0 & 0 & 51.6 & 51.6 & 1 & 51.6 & 0 \\
Weibull, $k=0.5$ & 5.00 & 86.7 & 86.7 & 1 & 86.7 & 0 & 74.4 & 74.4 & 1 & 74.4 & 0 \\
\hline
\end{tabular}
\end{table*}

We have the following observations.
\begin{itemize}
\item \emph{DFR service times.} For all five DFR distributions, the tuned parameters are $\alpha^*=1$ and $\tau^*=0$ for all $\lambda$, i.e., FP, PP and TP coincide. Preemption itself is highly beneficial (41\% to 87\% reduction in average AoI at $\lambda=5$),
but there is nothing to be gained from making update age-dependent decisions. This is consistent with the DFR property for which a fresh update is always stochastically shorter than the residual service time of the one in service, so it should always be preferred.
\item \emph{IFR service times.} The picture is reversed with FP in comparison to DFR service times. FP can perform very poorly because repeated preemptions can substantially delay successful service completions. The numerically tuned PP and TP policies provide limited improvements over NP for the IFR distributions considered. Particularly, gains are at most $7.1\%$ (Erlang-2) in average AoI and $9.9\%$ in average PAoI at $\lambda=5$, and zero for the deterministic and Weibull $k=3$ service times. For these two distributions, both tuned policies reduce to NP. For Erlang-4, PP coincides with NP, whereas TP yields small gains of $0.3\%$ in average AoI and $2.0\%$ in average PAoI at $\lambda=5$. In the IFR group, the gain in average AoI of TP is not monotone in $\lambda$ (Fig.~\ref{fig:gains}(a)), whereas the PAoI gain is.
\item \emph{Unimodal hazard rates.} This is the regime where age-dependent preemption pays off. To see this, for the lognormal distribution with $\sigma=0.75$ ($\mathrm{CV}^2=0.76$), the tuned TP policy reduces the average AoI by $18.8\%$ and the average PAoI by $15.8\%$ at $\lambda=5$, against $12.1\%$ and $7.9\%$ for tuned PP, while FP is $35\%$ worse than NP. For the log-logistic distribution, the gain is $13.7\%$ against $7.2\%$. The optimum thresholds are of the order of the mean service time ($\tau^*\approx0.8$ to $1.5$). As $\sigma$ increases, the lognormal hazard rate peaks earlier and the tail becomes heavier, so that the distribution behaves as a DFR distribution over the relevant range of ages: $\tau^*$ decreases to zero and the three preemptive policies merge, with large gains over NP ($71\%$ for $\sigma=1.25$). At the other end, for $\sigma=0.5$ ($\mathrm{CV}^2=0.28$) PP cannot improve upon NP at all, whereas TP still gains $2$--$3.5\%$.
\item \emph{TP versus PP.} In all 15 distributions and for all values of $\lambda$ we studied, the tuned TP policy performs at least as well as the tuned PP policy, with equality occurring, in particular, when both policies coincide with FP or NP. The advantage of TP is largest for unimodal hazard rates with moderate $\mathrm{CV}^2$, and the gap is wider for the average PAoI than for the average AoI. For several IFR and unimodal-hazard distributions, increasing $\lambda$ is associated with a decrease in $\alpha^*$ and an increase in $\tau^*$; this trend does not apply to the DFR group, where $\alpha^*=1$ and $\tau^*=0$ remain unchanged.
\item \emph{The role of $\mathrm{CV}^2$.} Within the IFR and unimodal groups, PP improves upon NP in terms of the average PAoI only when $\mathrm{CV}^2$ exceeds roughly $1/3$ (Weibull $k=1.5$, Erlang-2, lognormal $\sigma\ge0.75$, log-logistic), and not for the deterministic, Weibull $k=3$, Erlang-4 and lognormal $\sigma=0.5$ service times. TP provides gains over NP for several IFR and unimodal-hazard distributions, including cases in which PP offers no improvement (e.g., lognormal $\sigma=0.5$). However, TP also coincides with NP for Weibull $k=3$ and deterministic service times. A more detailed characterization of these numerical patterns, particularly for more general non-monotonic hazard-rate shapes, is left for future work.
\end{itemize}

\section{Numerical Example for Dual-source Case}\label{Multi-source}
The single-source results in Section~\ref{results_sec} show that the
benefits of threshold-based preemption depend strongly on the
service-time hazard rate. In particular, TP offers notable gains over
age-independent preemption policies for the unimodal-hazard distributions
considered. Motivated by these findings, we investigate a dual-source
($C=2$) self-preemptive status update system with lognormal service
times. Unlike in the single-source case, the preemption parameters of
one source also affect the freshness of the other through their shared
server. We therefore examine the policies under joint per-source tuning.

We compare the
four packet management policies introduced in Section~\ref{system_sec}:
the NP policy, 
the FP policy,
the PP policy with per-source preemption
probabilities $(\alpha_1,\alpha_2)$, and the proposed 
TP policy with per-source thresholds $(\tau_1,\tau_2)$.
In this example, the service times of both sources follow lognormal
distributions with unit mean, $\mathbb{E}[S_c]=1$, obtained by setting
   $ \mu_c=-\frac{\sigma_c^2}{2},~ c\in\{1,2\}.$
We set $\sigma_1=\sigma_2=1$, a unimodal-hazard case in which
Section~\ref{results_sec} showed benefits from age-dependent
preemption. This setting allows us to examine how these benefits
persist when two sources compete for a shared server.

\subsubsection{Policy-Parameter Tuning via Grid Search and
Nelder--Mead Refinement}
\label{Nelder--Mead Refinement}
For the PP and TP policies, the per-source policy parameters are jointly
tuned to minimize either the sum average AoI
    $\sum_{c=1}^{C}\mathbb{E}[\Delta_c],$
or the sum average PAoI
    $\sum_{c=1}^{C}\mathbb{E}[\Phi_c],$
depending on the performance metric under consideration. The average
AoI and PAoI are evaluated using Theorems~\ref{thm:paoi}
and~\ref{thm:aoi}. Their different dependence on the inter-delivery
time can also be seen from \eqref{meanPAoIexp} and
\eqref{meanAoIexp_O}: the average PAoI depends on
$\mathbb{E}[V_c]$, whereas the average AoI additionally depends on
the second moment $\mathbb{E}[V_c^2]$.

For PP, the parameter space is
$(\alpha_1,\alpha_2)\in[0,1]^2$, whereas for TP the thresholds are
searched over $(\tau_1,\tau_2)\in[0,\tau_{\max}]^2$. We employ a
two-stage numerical tuning procedure. In the first stage, a coarse grid
search evaluates the objective over the Cartesian product of uniformly
spaced candidate values in the two parameter dimensions. For PP, we use
$21$ equally spaced values in $[0,1]$ for each preemption probability,
resulting in $21^2$ candidate pairs. For TP, we set
$\tau_{\max}=5$ and use $31$ equally spaced values in
$[0,\tau_{\max}]$ for each threshold, resulting in $31^2$ candidate
pairs.

The best grid point is then used as the initial point for a local
Nelder--Mead refinement~\cite{Nelder_Mead}. The Nelder--Mead simplex
method is implemented using MATLAB's \texttt{fminsearch} routine. Since \texttt{fminsearch} is
unconstrained, candidate points outside the feasible parameter
region are penalized in the objective function. For
$C=2$, the simplex consists of three candidate parameter pairs. The
stopping tolerances for the parameter vector and objective function are
set to $10^{-4}$ and $10^{-5}$, respectively. The maximum numbers of
iterations and objective-function evaluations are set to $500$ and
$1500$, respectively.

For every candidate policy parameter pair, the absorption probabilities
in \eqref{prob_preempt} and \eqref{prob_complete}, together with the
conditional lifetime moments characterized by
Lemma~\ref{lemma:condpdf}, are numerically evaluated. These quantities
are then used in the inter-delivery-time expressions
\eqref{meanVc}--\eqref{secondmomentVc} and, finally, in the AoI and
PAoI expressions of Theorems~\ref{thm:paoi} and~\ref{thm:aoi}.
The numerical integrations are performed after the change of variable
$y=\frac{\log t-\mu_c}{\sigma_c},$
using a uniform grid over $[-10,10]$ with $20001$ points.

FP is a boundary case of both optimized policy classes. In particular, PP reduces to FP for
$(\alpha_1,\alpha_2)=(1,1)$, whereas TP reduces to FP for $(\tau_1,\tau_2)=(0,0)$. These boundary points are explicitly compared with the refined solutions when selecting the final numerical solution. We denote the resulting numerically tuned policies by PP$^\star$ and
TP$^\star$ as before. The PP and TP policies are retuned separately for the AoI
and PAoI experiments, so PP$^\star$ and TP$^\star$ are numerically tuned for the metric
displayed in the corresponding figure.

\subsubsection{Sum Average AoI versus the Source Arrival Rates}

\begin{figure*}[!t]
\centering
\subfigure[$\lambda_2=0.25$]{
\includegraphics[width=0.43\textwidth]{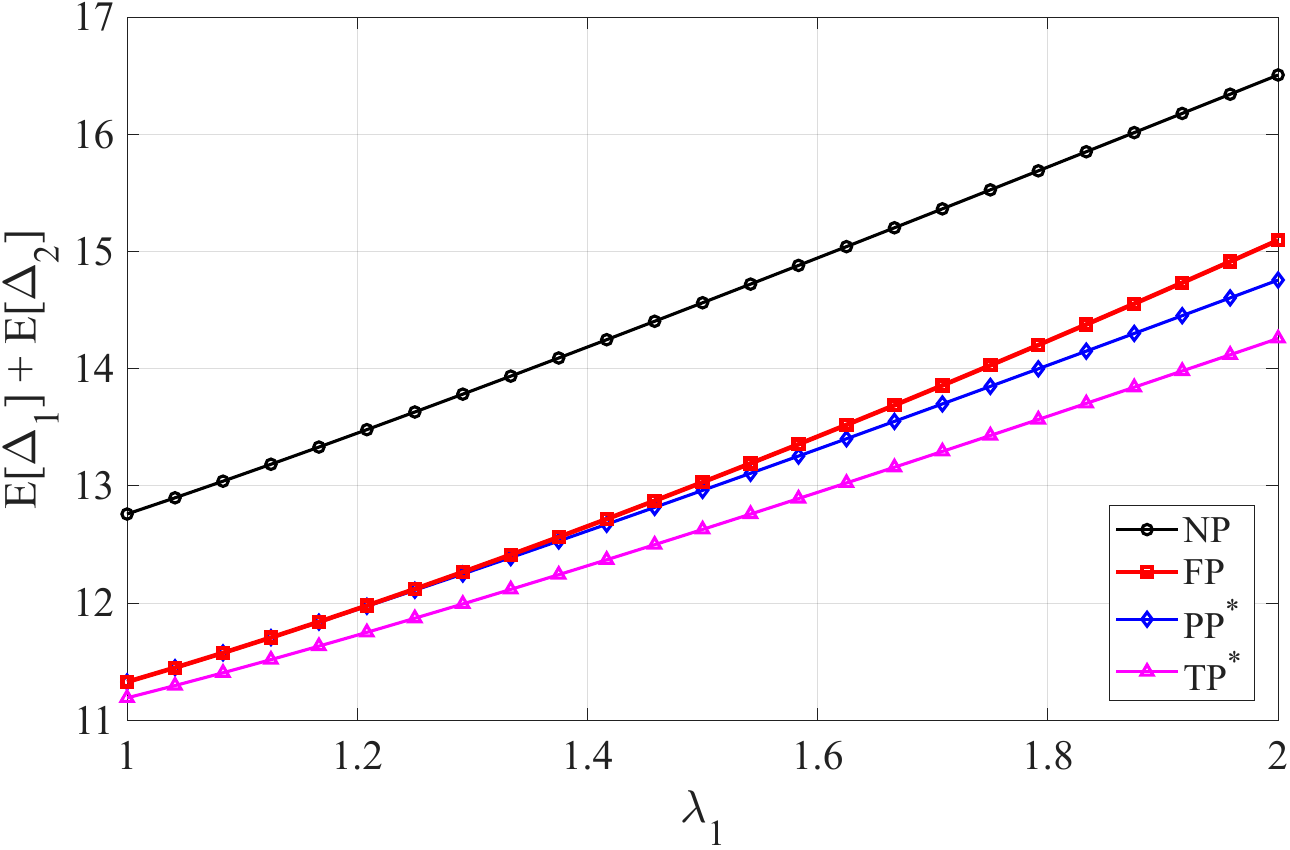}
\label{Lam2_p25}
}
\subfigure[$\lambda_2=1$]
{
\includegraphics[width=0.43\textwidth]{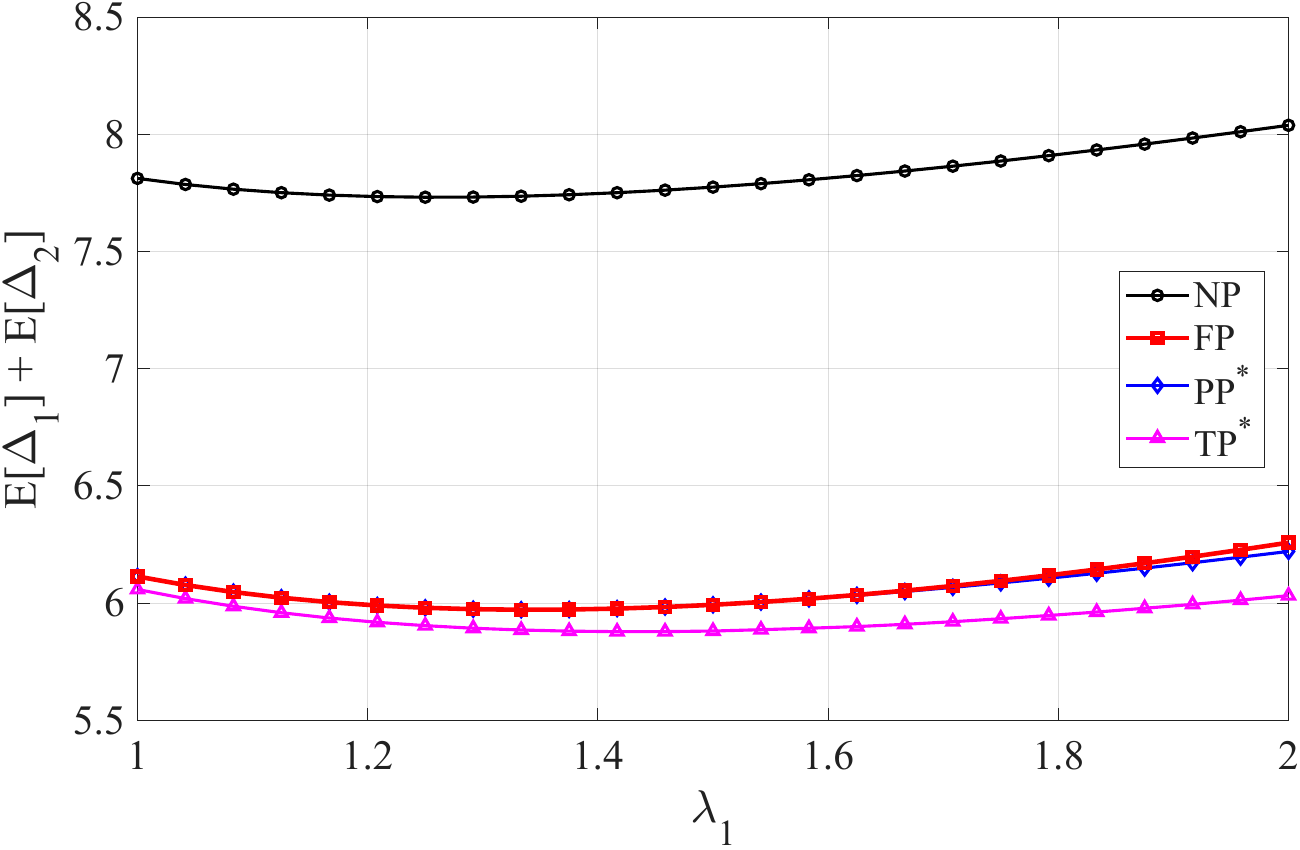}
\label{Lam2_1}
}
\subfigure[$\lambda_2=2$]{
\includegraphics[width=0.43\textwidth]{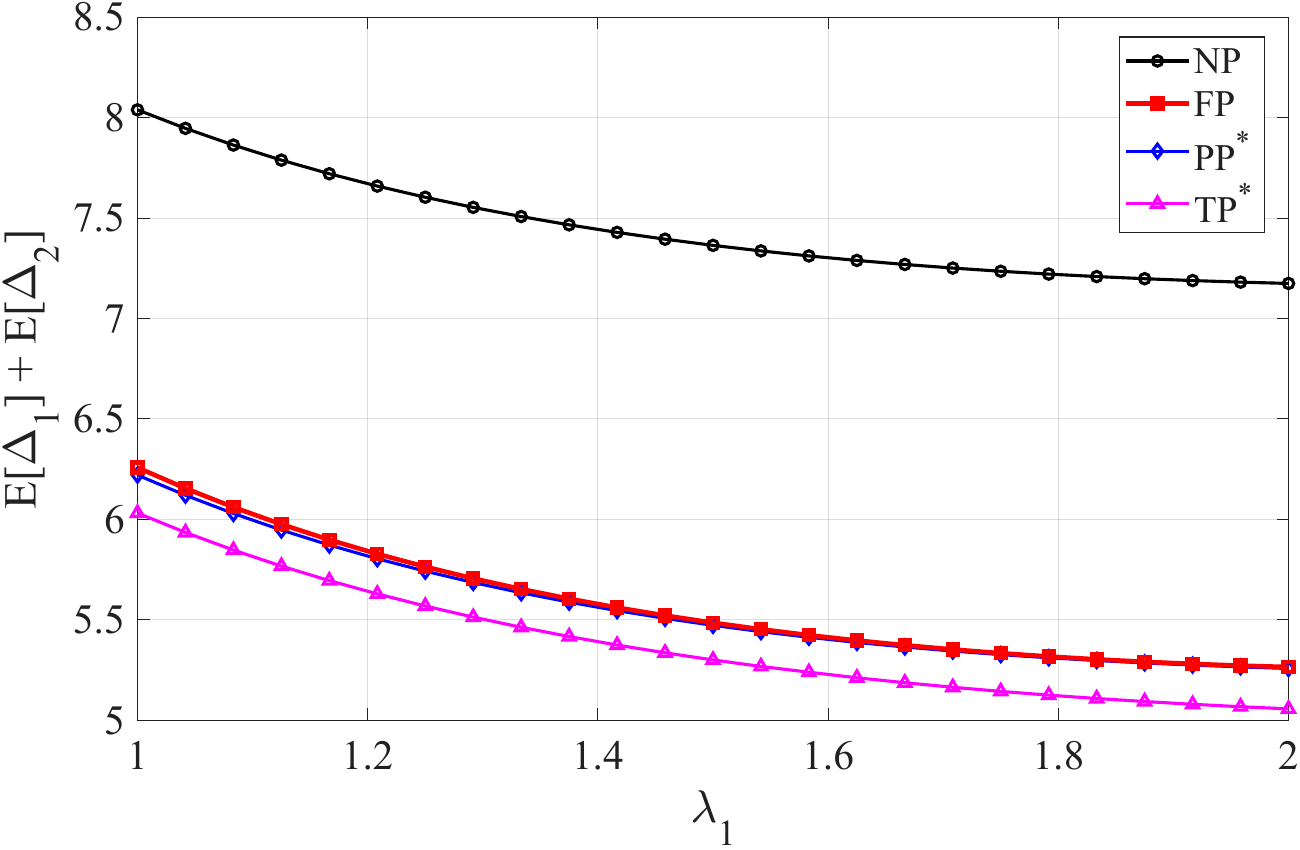}
\label{Lam2_2}
}
\caption{Sum average AoI under NP, FP, PP$^\star$, and TP$^\star$
as a function of $\lambda_1$ for
$\sigma_1=\sigma_2=1$ and unit-mean lognormal service times:
(a) $\lambda_2=0.25$, (b) $\lambda_2=1$, and
(c) $\lambda_2=2$. The parameters of PP$^\star$ and TP$^\star$
are jointly tuned to minimize
$\mathbb{E}[\Delta_1]+\mathbb{E}[\Delta_2]$ at each arrival-rate pair.}
\label{fig:sum_aoi_lambda1}
\end{figure*}

Fig.~\ref{fig:sum_aoi_lambda1} shows the sum average AoI
$\mathbb{E}[\Delta_1]+\mathbb{E}[\Delta_2]$ under NP, FP,
PP$^\star$, and TP$^\star$ as a function of $\lambda_1$. We set
$\sigma_1=\sigma_2=1$ and consider
$\lambda_2\in\{0.25,1,2\}$. At every arrival-rate pair
$(\lambda_1,\lambda_2)$, PP$^\star$ and TP$^\star$ are retuned to
minimize the sum average AoI.

When $\lambda_2=0.25$, Fig.~\ref{Lam2_p25} shows that the sum average
AoI increases with $\lambda_1$. In this regime, source~2 generates
updates relatively infrequently. Increasing $\lambda_1$ allows
source~1 to capture the idle server more frequently and increases the
waiting time of source~2 through the cross-source occupancy term in
\eqref{eta0c}. The resulting degradation in the freshness of source~2
dominates the improvement experienced by source~1.

For $\lambda_2=1$, Fig.~\ref{Lam2_1} shows a non-monotone behavior.
The sum average AoI initially decreases with $\lambda_1$, reaches a
minimum at an intermediate arrival rate, and then increases. Increasing
$\lambda_1$ makes fresh source~1 updates available more frequently,
which tends to improve its freshness. At the same time, the increased
server occupancy associated with source~1 increases the cross-source
penalty experienced by source~2. At smaller $\lambda_1$, the former
effect dominates; as $\lambda_1$ increases, the two effects balance and
eventually the cross-source penalty becomes dominant.

For $\lambda_2=2$, Fig.~\ref{Lam2_2} shows a decreasing sum average
AoI over the considered range. Since source~2 initially generates
updates more frequently than source~1, increasing $\lambda_1$ reduces
the imbalance between the source arrival rates. The freshness
improvement associated with the increased source~1 update rate dominates
the additional cross-source penalty, resulting in a reduction in the
sum average AoI. The different trends in
Figs.~\ref{Lam2_p25}--\ref{Lam2_2} illustrate that, although
preemption is self-preemptive, the freshness of the two sources remains
coupled through their shared use of the server.

Across all considered cases, TP$^\star$ achieves the smallest sum
average AoI among the compared policies. NP cannot interrupt unusually long service realizations,
which can significantly increase the inter-delivery time and, in
particular, its second moment in \eqref{meanAoIexp_O}. FP avoids such
long service episodes but may unnecessarily discard recently started
updates. PP$^\star$ provides additional flexibility through a fixed
preemption probability, but its decision remains independent of the
elapsed service time. In contrast, TP$^\star$ preserves recently
started updates and allows preemption only after the elapsed service
time exceeds the optimized threshold. For the lognormal service
distribution, this enables TP$^\star$ to limit long service episodes
while avoiding overly aggressive preemption, yielding the smallest sum
average AoI among the compared policies.

\subsubsection{Sum Average PAoI versus the Source Arrival Rates}

\begin{figure*}[!t]
\centering
\subfigure[$\lambda_2=0.25$]{
\includegraphics[width=0.43\textwidth]{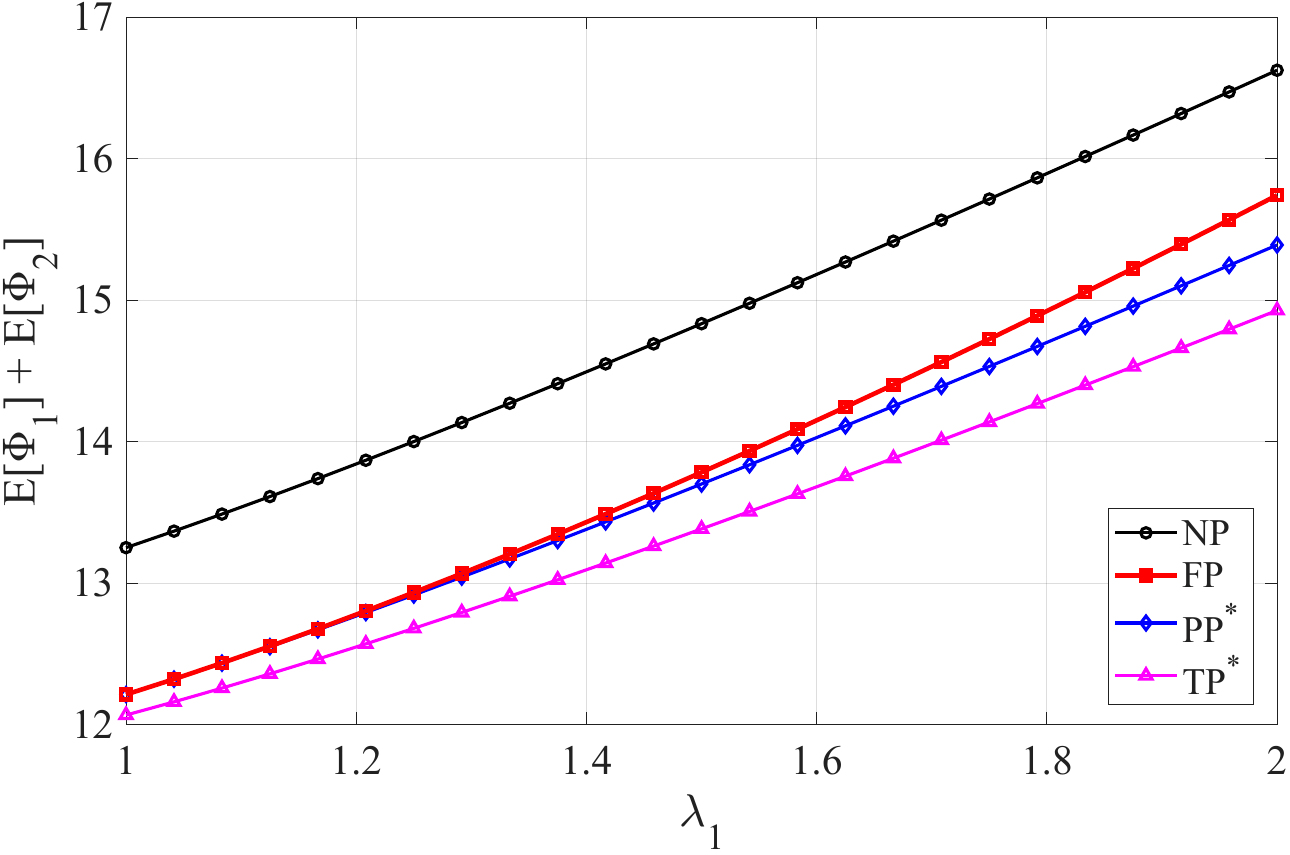}
\label{Paoi_Lam2_p25}
}
\subfigure[$\lambda_2=1$]
{
\includegraphics[width=0.43\textwidth]{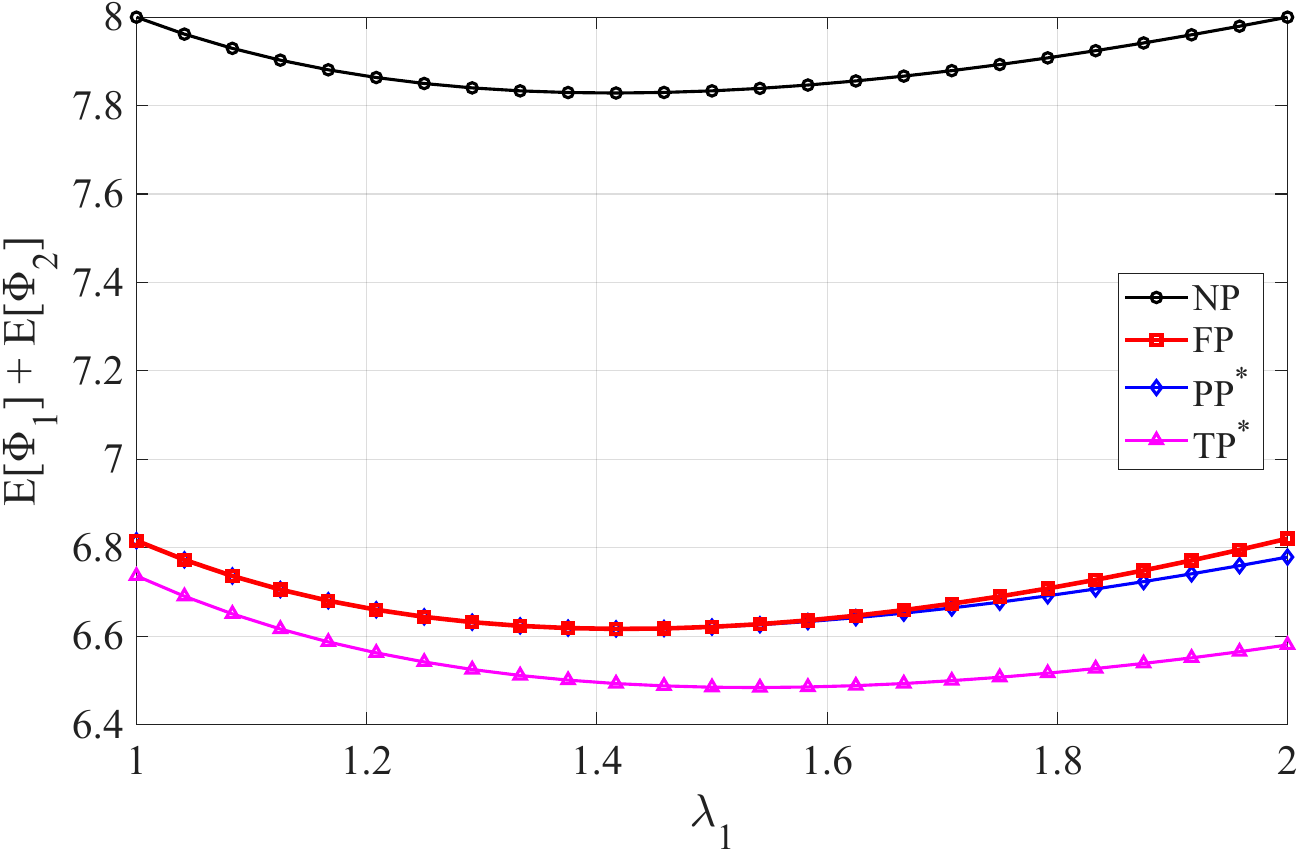}
\label{Paoi_Lam2_1}
}
\subfigure[$\lambda_2=2$]{
\includegraphics[width=0.43\textwidth]{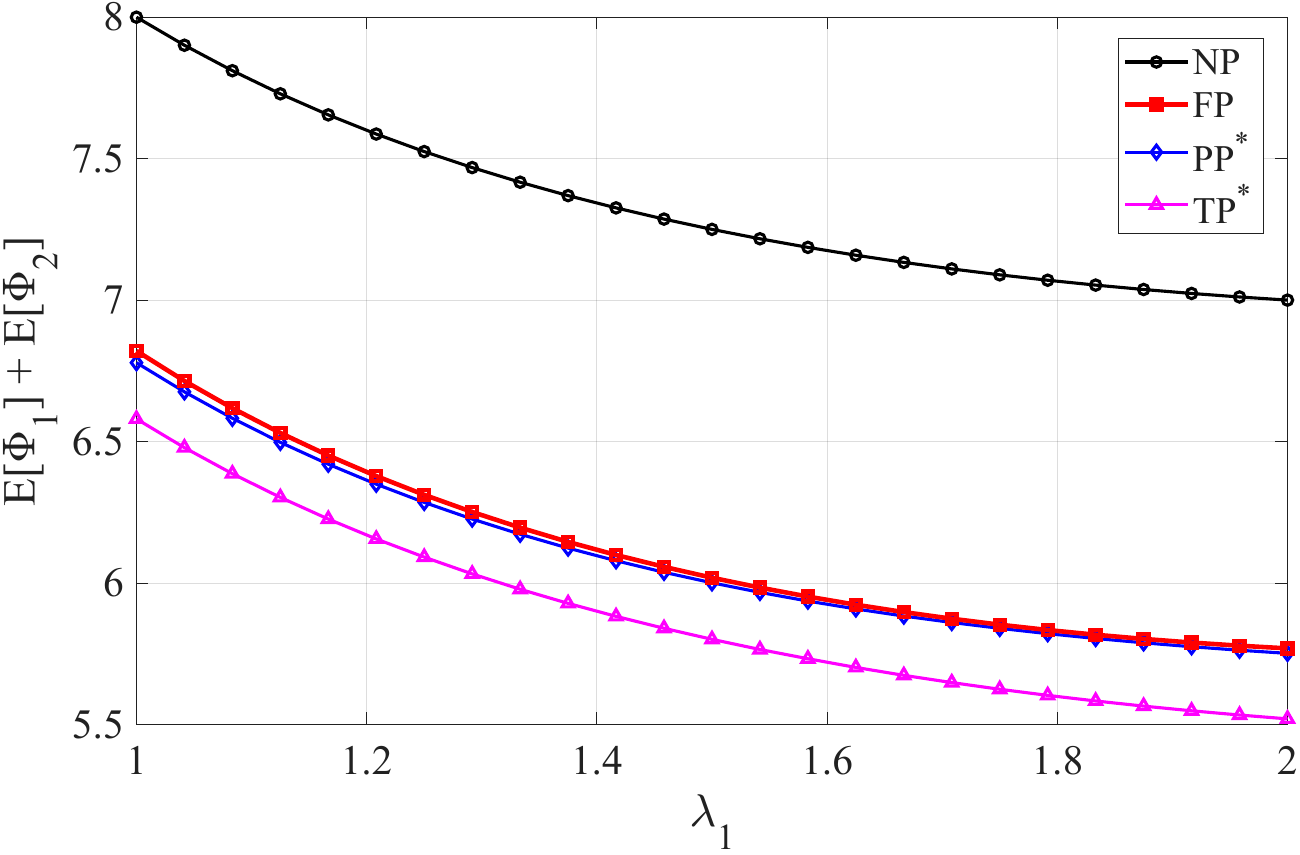}
\label{Paoi_Lam2_2}
}
\caption{Sum average PAoI under NP, FP, PP$^\star$, and TP$^\star$
as a function of $\lambda_1$ for
$\sigma_1=\sigma_2=1$ and unit-mean lognormal service times:
(a) $\lambda_2=0.25$, (b) $\lambda_2=1$, and
(c) $\lambda_2=2$. The parameters of PP$^\star$ and TP$^\star$
are jointly tuned to minimize
$\mathbb{E}[\Phi_1]+\mathbb{E}[\Phi_2]$ at each arrival-rate pair.}
\label{fig:sum_paoi_lambda1}
\vspace{-2mm}
\end{figure*}

Fig.~\ref{fig:sum_paoi_lambda1} shows the sum average PAoI
$\mathbb{E}[\Phi_1]+\mathbb{E}[\Phi_2]$ under the same system
parameters considered in Fig.~\ref{fig:sum_aoi_lambda1}. For each
arrival-rate pair, PP$^\star$ and TP$^\star$ are retuned to minimize
the sum average PAoI.

The PAoI results exhibit trends similar to those observed for the
average AoI. For $\lambda_2=0.25$, Fig.~\ref{Paoi_Lam2_p25} shows an
increase in the sum PAoI with $\lambda_1$, since the additional
source~1 occupancy increases the mean waiting and inter-delivery times
of the low-rate source~2. For $\lambda_2=1$,
Fig.~\ref{Paoi_Lam2_1} exhibits a minimum at an intermediate
$\lambda_1$, reflecting the balance between the improved delivery
frequency of source~1 and the increased cross-source penalty imposed on
source~2. When $\lambda_2=2$, Fig.~\ref{Paoi_Lam2_2} shows a decreasing
sum PAoI as increasing $\lambda_1$ reduces the arrival-rate imbalance
between the two sources.

The TP$^\star$ policy again achieves the smallest sum average PAoI among the compared policies in all three cases.
Since the average PAoI in \eqref{meanPAoIexp} depends on
$\mathbb{E}[U_c]$ and $\mathbb{E}[V_c]$, the gain of TP$^\star$ is not
only a consequence of reducing the second moment of the inter-delivery
time. The results show that elapsed-service-time-dependent preemption
can also improve the mean delivery-cycle behavior compared with the
age-independent NP, FP, and PP policies.

Overall, the AoI and PAoI results emphasize the importance of joint
per-source tuning. According to \eqref{eta0c} and
\eqref{eta0csecond}, the inter-delivery statistics of source $c$ depend
on the service episodes of the competing source. Consequently, changing
$\tau_j$ or $\alpha_j$ affects not only the freshness of source $j$,
but also the waiting-time statistics and freshness of the other source.
Thus, tuning each source in isolation need not minimize the sum
AoI or PAoI, motivating joint per-source optimization in multi-source systems.

\section{Conclusions}\label{conclusions_sec}
\vspace{-1mm}
We studied the average AoI and PAoI in a status update system with an arbitrary number of independent sources sharing a single server, generally distributed service times, and age-dependent self-preemption probabilities. Using a non-homogeneous absorbing CTMC for the lifetime analysis and a semi-Markov chain for the inter-delivery time characterization, we obtained analytical expressions for the average AoI and PAoI of each source under general per-source preemption rules, with the non-preemptive, fully self-preemptive, and probabilistically preemptive policies emerging as special cases.

Numerical results for fifteen service-time distributions in the single-source setting demonstrate that the benefits of preemption depend strongly on the shape of the service-time hazard rate. For the DFR distributions considered, fully self-preemptive operation is particularly effective, whereas the IFR distributions generally favor limited or no preemption. Threshold-based preemption provides its most notable additional benefits over age-independent policies for distributions with unimodal hazard rates. For a dual-source system with lognormal service times, the jointly tuned threshold-based policy achieves lower sum average AoI and PAoI than the considered age-independent policies. The results also reveal cross-source interactions: the optimized preemption threshold for each source depends not only on its own arrival and service statistics but also on the competing load from the other source, highlighting the importance of joint per-source tuning in multi-source systems.

\appendices

 \section{Proof of Lemma~\ref{lemma:condpdf}}\label{proof-lemma:condpdf}
 \vspace{-1mm}
 We derive \eqref{fTpc}; the derivation of \eqref{fTqc} follows by identical arguments with $\lambda_c p_c(x)$ replaced by $q_c(x)$.

Let $f_T(x)$ denote the unconditional pdf of the lifetime $T$, i.e., the pdf of the absorption time of $Y(x)$ regardless of which absorbing state is entered.
Since absorption from the transient state $0$ occurs at rate $\lambda_c p_c(x) + q_c(x)$ when the elapsed time is $x$, and the probability of still being in state $0$ at time $x$ is $\pi_c(x)$, the unconditional pdf of $T$ is given as
\begin{align}\label{fT_unconditional}
    f_T(x) & 
    = {\dv{x}(1-\pi_c(x))} 
    \\&\nonumber 
 =\pi_c(x)\bigl(\lambda_c p_c(x) + q_c(x)\bigr). 
\end{align}
By Bayes' theorem, the conditional pdf of $T$ given the event $Y_1 = \{Y(\infty)=1\}$ is given as
\begin{align}\label{bayes_step}
    f_{T_{p_c}}(x)
    &= f_{T \mid Y_1}(x)
    \\&\nonumber
    = \frac{f_T(x)\,\Pr(Y_1 \mid T = x)}{\Pr(Y_1)}. 
\end{align}
Given that absorption occurs exactly at time $T = x$, the probability that it is a preemption event (rather than a completion) equals the ratio of the preemption rate to the total absorption rate at time $x$, which is given as
\begin{align}
    \Pr(Y_1 \mid T = x) &= \dfrac{\lambda_c p_c(x)}{\lambda_c p_c(x) + q_c(x)}. \label{cond_preempt_given_T}
\end{align}
Substituting \eqref{fT_unconditional} and \eqref{cond_preempt_given_T} into \eqref{bayes_step}, and using $\Pr(Y_1) = \varepsilon_c^{p}$, establishes \eqref{fTpc}. The derivation of \eqref{fTqc} follows identically with $\Pr(Y_2 \mid T=x) = q_c(x)/(\lambda_c p_c(x) + q_c(x))$ and $\Pr(Y_2) = \varepsilon_c^{q}$.

\bibliographystyle{IEEEtran}
\bibliography{ Nbiblmultisource_consolidated.bib}
\end{document}